\documentclass[authoryear,review,12pt]{elsarticle}
\usepackage[utf8]{inputenc}
\usepackage[titletoc,title]{appendix}
\usepackage[utf8]{inputenc}
\usepackage[T1]{fontenc}
\usepackage{mathtools}
\usepackage[thinc]{esdiff}
\usepackage[dvipsnames]{xcolor}
\usepackage{wrapfig}
\usepackage{graphicx}
\usepackage{subcaption}
\usepackage{setspace}
\usepackage[hidelinks]{hyperref} 
\usepackage{amsmath}

\usepackage{amsmath,amsfonts,amssymb,mathtools}

\usepackage{graphicx,float}

\usepackage{multirow}

\usepackage[ruled,vlined]{algorithm2e}
\usepackage{algorithmic}

\usepackage{xcolor-material}
\usepackage{soul}

\newcommand{\beginsupplement}{%
  \setcounter{table}{0}
  \setcounter{figure}{0}
  \renewcommand{\thetable}{S\arabic{table}}%
  \renewcommand{\thefigure}{S\arabic{figure}}%
  \renewcommand{\theHtable}{Supplement.\thetable}%
  \renewcommand{\theHfigure}{Supplement.\thefigure}     
}

\newcommand{\citeg}[1]{\citep[e.g.,][]{#1}}

\journal{EPSL}
\begin{document}
\begin{frontmatter}
\pagenumbering{arabic} 
\title{Deep learning for laboratory earthquake prediction and autoregressive forecasting of fault zone stress}

\author[sapienza]{Laura Laurenti\corref{cor1}}
\author[sapienza]{Elisa Tinti}
\author[sapienza]{Fabio Galasso}
\author[sapienza]{Luca Franco}
\author[sapienza,penn]{Chris Marone}

\address[sapienza]{Università  La Sapienza, Rome, Italy}

\address[penn]{The Pennsylvania State University, University Park, PA, USA}
\cortext[cor1]{laura.laurenti@uniroma1.it}


\begin{abstract}

Earthquake forecasting and prediction have long and in some cases sordid histories but recent work has rekindled interest based on advances in early warning, hazard assessment for induced seismicity and successful prediction of laboratory earthquakes. 
In the lab, frictional stick-slip events provide an analog for earthquakes and the seismic cycle.
Labquakes are also ideal targets for machine learning (ML) because they can be produced in long sequences under controlled conditions. Indeed, recent works show that ML can predict several aspects of labquakes using fault zone acoustic emissions (AE). 
Here, we extend these works with: 1) deep learning (DL) methods for labquake prediction, 2) by introducing an autoregressive (AR) forecasting DL method to predict fault zone shear stress, and 3) by expanding the range of lab fault zones studied. The AR methods allow forecasting stress at future times via iterative predictions using previous measurements.
Our DL methods outperform existing ML models and can predict based on limited training. We also explore forecasts beyond a single seismic cycle for aperiodic failure. We describe significant improvements to existing methods of labquake prediction and  demonstrate: 1) that DL models based on Long-Short Term Memory and Convolution Neural Networks predict labquakes under conditions including pre-seismic creep, aperiodic events and alternating slow/fast events and 2) that fault zone stress can be predicted with fidelity, confirming that acoustic energy is a fingerprint of fault zone stress. Our DL methods predict time to start of failure (TTsF) and time to the end of Failure (TTeF) for labquakes. Interestingly, TTeF is successfully predicted in all seismic cycles, while the TTsF prediction varies with the amount of preseismic fault creep. 
We report AR methods to forecast the evolution of fault stress using three sequence modeling frameworks: LSTM, Temporal Convolution Network and Transformer Network. AR forecasting is distinct from existing predictive models, which predict only a target variable at a specific time.
The results for forecasting beyond a single seismic cycle are limited but encouraging. 
Our ML/DL models outperform the state-of-the-art and our autoregressive model represents a novel framework that could enhance current methods of earthquake forecasting.\\

\end{abstract}

\bigskip

\begin{keyword}


Machine learning, Neural Networks,  laboratory earthquakes, earthquake prediction, earthquake forecasting, auto-regressive forecasting, LSTM, TCN, Transformer

\end{keyword}

\end{frontmatter}



\section{Introduction}

Earthquake forecasting and prediction have long been of interest because of the obvious practical and societal implications. While research has waxed and waned with many failed directions, recent work on early warning systems, hazard assessment, and precursors has provided renewed interest 
\citep{ref:Allen2022,ref:Ben-Zion2002,ref:Beroza2021,ref:Wang2021, ref:Denolle2014,ref:Kohler2020,ref:Kong2019,ref:Pritchard2010}.  Laboratory work has fueled this interest via: 1) the discovery that machine-learning can predict several aspects of lab earthquakes \citep{ref:RouetLeduc2017,ref:RouetLeduc2018,ref:Johnson2021} and 2) recent work on the mechanisms of precursors to labquakes that adds to earlier studies  \citep{ref:Shreedharan2021, ref:Bolton2021, ref:Dieterich1978, ref:Scholz1968, ref:Dresen2020,ref:Scuderi2016,ref:Hedayat2014,ref:Acosta2019,ref:Johnson2013,ref:Main1989,ref:Main1992,ref:Thompson2005,ref:Passelegue2017}. 

Recent works show that acoustic emissions can be used to predict labquake failure time, fault zone shear stress, and labquake stress drop \citep{ref:RouetLeduc2017, ref:RouetLeduc2018, ref:Lubbers2018, ref:Hulbert2019, ref:Bolton2020}. Existing works also show that seismic radiation from lab faults scales with time to failure and that its recent history can be used to predict the current fault zone stress state \citep{ref:Bolton2019,ref:Corbi2019,ref:Jaspereson2021} with both active and passive seismic monitoring of the fault zone \citep{ref:Shreedharan2021, ref:Shokouhi2021}. 
These studies show that both passive signals coming from the fault zone and active acoustic signals passing through the fault can be used to predict failure. The active source signals record changes in fault properties prior to failure and thus offer the possibility of incorporating physics-based models (i.e, of asperity contact mechanics) in the ML/DL algorithm. Other studies show that ML can be used to connect lab AE events with fault zone microstructure \citep{ref:Chaipornkaew2022,ref:Trugman2020,ref:Ma2022} and that ML methods can be augmented by directly incorporating physics into the prediction models \citep{ref:Raissi2019}. Here, we extend this approach to investigate the use of DL methods and also to introduce a new approach based on autoregressive forecasting. The AR methods are distinct because they predict future time horizons using present and past values rather than the current time based on previous values. 

Recent lab studies have also identified reliable precursors to failure in connection with labquake prediction. These include strong relations between acoustic transmissivity, elastic properties and fault
strength \citep{ref:Schubnel2006, ref:Nagata2008, ref:Scuderi2016, ref:Tinti2016, ref:Shreedharan2020, ref:Shreedharan2021, ref:Bolton2021, ref:Riviere2018}. 
Other previous studies provided insight on labquake nucleation processes and the evolution of frequency-magnitude (b-value) statistics during the seismic-cycle prior to failure \citep{ref:Dresen2020, ref:Riviere2018, ref:Scholz2015,ref:Goebel2017,ref:Kwiatek2014,ref:Latour2013,ref:Shreedharan2020,ref:Johnson2021,ref:McBeck2020,ref:McLaskey2014}.
These studies provide a framework for understanding rupture nucleation and therefore how machine learning can predict labquakes using statistics of the continuous AE signal emanating from lab faults \citep{ref:Hulbert2019}. 
Previous works have used ML to predict geodetic signals of episodic tremor and slip and other data from tectonic faults \citep{ref:Hulbert2020,ref:Johnson&Johnson2021,ref:Wang2021, ref:McBrearty2019}.

The majority of existing ML studies of labquakes use decision-tree-based algorithms and a gradient boosting framework (e.g., XGBoost model). In such studies models are built and errors are minimized by gradient descent methods using multiple learning algorithms that iteratively improve predictive performance beyond that of a single classical learning algorithm.
These studies show that it is possible to successfully predict labquakes with reasonable accuracy. The success of the original approaches was based on continuous records of AE events broadcast from the lab fault zones. These AE represent a form of lab microearthquakes. AE are also detected during stable frictional slip. In this case, their origin is less clear, as they could represent microfracture events or micro-instabilities that do not impact the macroscopic strength (and therefore they do not appear as a stress drop). 

Despite the dramatic recent advances in lab earthquake prediction our understanding of how and why these methods work is far from complete. Simple questions remain such as how the AE signal scales with labquake timing and magnitude and how AE signal characteristics encode fault zone shear stress even during aperiodic failure.  
Here, we address these questions by exploring a wider range of ML/DL methods. We also address the open discussion regarding the predictability of earthquakes and if faults slip deterministically or stochastically. 

We explore two different problems and three families of Neural Network (NN) architectures. The two problems refer to two different tasks: prediction and forecasting. The former has 
been addressed in previous works \citep{ref:Hulbert2019,ref:RouetLeduc2017}, and the main goal is how to predict the present value of target variables (e.g., shear stress) given past information and some memory of sequence evolution. 
With the second problem we introduce a novel autoregressive (AR) forecasting approach to predict not only the present value of the target variable, but also its step-by-step future evolution. 

The three families of NN architectures adopted in this paper are Long short-term memory (LSTM), Temporal Convolutional Network (TCN), and Transformer Network (TF).
 LSTM is a Recurrent Neural Network (RNN) designed to capture long-range data dependencies in time series and sequential data \citep{ref:Hochreiter1997}. RNNs are 
Deep Neural Networks (DNNs) that recursively update an internal status (this may represent the current status of the fault) from which predictions are made.  This approach provides the capability of modeling sequences of events. LSTM is designed to solve gradient vanishing problems of RNN's for long-term prediction.
LSTM has shown great potential for modeling seismic time series \citep{ref:Johnson2021}.
TCN, one of the simplest NN architectures for sequence modeling, is a Convolutional Neural Network (CNN) with adjustments that allow longer sequences by addressing the relation between the depth and the length of the considered input sequence.
Because it involves convolution, TCN may be applied to sequences of variable length. 
The TF architecture is the most recent and it has shown promising results for earthquake detection \citep{ref:Mousavi2020}. It's based on mechanisms of self-attention and is effective in capturing long-term dependencies in a variety of sequence modeling tasks \citep{ref:Vaswani2017}. 
One notable weakness of TF is its complexity and that it requires significantly larger datasets to be trained than other architectures.

\section{Laboratory Earthquake Experiments}

\begin{figure}[H]
    \includegraphics[width=0.95\textwidth]{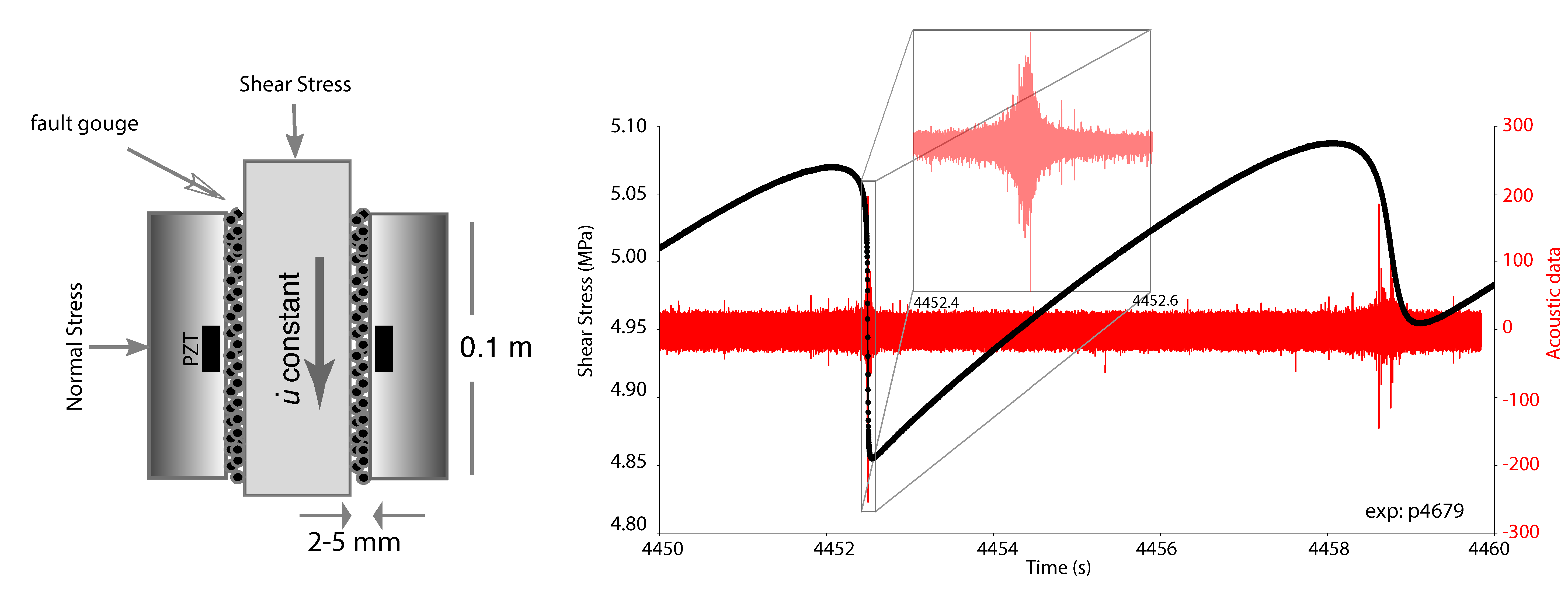}
    \caption{Left: schematic of the double direct shear configuration used for the experiments. Granular layers simulate an earthquake fault and are sheared between rough surfaces. Acoustic signals are recorded with Piezo-Electric Transducers (PZTs) embedded in the loading blocks. Faults are loaded with a normal stress that is maintained constant via servo control. The central block is driven downward at constant displacement rate to produce frictional shear. Acoustic emission (AE) data are recorded continuously at a sampling rate of 4 MHz. 
    Right: typical data for shear stress and AE coming from the fault zone. Zoomed window above shows data from a lab earthquake. Note that the acoustic signal differs for the large labquake compared to the smaller event that follows (near 4459 s).}
    \label{fig:pict}
\end{figure}

We use data from experiments conducted in the double-direct shear (DDS) configuration with a biaxial deformation machine (see Figure \ref{fig:pict}). The DDS experiments consist of two identical fault zones sheared simultaneously between three loading blocks. Two stationary side blocks are supported from below and a central block is driven between them to impose shear. 

Our experiments were performed following a standard procedure that was developed to obtain highly reproducible results \citep{ref:Karner1998,ref:Bolton2020}. First, fault normal stress was imposed with a servo-controlled hydraulic system and maintained at a controlled value throughout shear. Then, fault zone shear was imposed by driving the central block of the DDS configuration at a constant displacement rate, thus imposing a constant shear strain rate within the fault zone (Table \ref{tab:exp results pt1}). Fault zones were composed of granular materials that began as 3-mm thick layers of nominal contact area 10 × 10 $cm^2$.  
We use data from two types of lab fault zones: 1) glass beads (experiment p4581) and 2) quartz powder, (experiments p4679 and p5198). Table \ref{tab:exp results pt1} provides experiment details. The glass beads range in size from 100-150 µm and for these experiments the normal stress is held below 10 MPa so that grain fracturing is minimal  \citeg{ref:Mair2002}. The quartz powder has a mean particle size of 10 µm \citep{ref:Bolton2020} and a power-law particle size distribution that does not change significantly with shear for the stresses of our experiments. The fault normal stress was 7 MPa for experiment p4679 and it ranged from 6 to 11 MPa in experiment p5198 and from 2 to 8 MPa in p4581.  These experiments exhibit a range of stick-slip behaviors from fast to slow \citep{ref:Leeman2016} and have seismic cycles that range from highly periodic to complex and aperiodic.

We study a range of frictional sliding behaviors from stable sliding to highly unstable slip. In the framework of rate-and-state friction this range of behaviors is understood in terms of the ratio of loading stiffness to a critical rate of fault weakening with slip, which scales with normal stress \citep{ref:Leeman2016}. 
Each experiment includes hundreds of lab earthquakes (Figure \ref{fig:ALLexp}).  
By adjusting the loading stiffness and/or the fault normal stress, the same fault gouge can host slow-slip events, fast lab earthquakes, or complex, quasi-periodically events
\citep{ref:Leeman2016, ref:Scuderi2017, ref:Veedu2020}. 
Mechanical data of stress and strain are recorded at 1 kHz and acoustic signals are recorded at 4 MHz using  lead-zircon-titanate piezoceramic sensors embedded in the DDS (Figure \ref{fig:pict}) \citep{ref:Riviere2018,ref:Bolton2020}.  A high-precision signal is used to synchronize the mechanical and acoustic data. 

In the lab three stages of the seismic cycle can be identified: an initial --interseismic-- stage of stress increase, a stage with pre-seismic slip and then a last stage with a co-seismic stress drop (Figures \ref{fig:pict} and \ref{fig:Signals' shapes}). The interseismic period of the lab seismic cycle is identified from the initial, linear-elastic portion of the curve, where load increases in proportion to stiffness. Pre-seismic slip is marked by a deviation from elastic loading. This typically corresponds to a relatively short time interval preceding failure and seismic energy release. Pre-seismic slip is part of the labquake nucleation process. 
We study data from three experiments chosen to represent a range of lab earthquake behaviours. In some cases the events are quasi-periodic, as for experiment p5198 (Figure \ref{fig:Signals' shapes} a), and in others, as for experiment p4679 (Figure \ref{fig:Signals' shapes} d) they are aperiodic with alternating slower and faster events. 
Experiments like p4679 often have complex stress evolution throughout the lab seismic cycle. Some of this complexity can be seen at the scale of the whole experiment (Figure \ref{fig:ALLexp}) and other aspects of it are visible only at a larger scale (Figure \ref{fig:Signals' shapes}). 
In the case of p4581, the pre-seismic phase of the seismic cycle is characterized by almost no pre-slip prior to failure.
The acoustic emission record for this type of experiment is distinct because the signal variance is quite low until just prior to the labquake (Figure \ref{fig:Signals' shapes p4581}). Nonetheless there is still AE activity before the main labquake. In contrast, experiment p5198 shows significant pre-seismic slip before the peak stress and failure (see Figure \ref{fig:Signals' shapes}). Both of these experiments show somewhat regular, quasi-periodic seismic cycles (Figures \ref{fig:Signals' shapes} and \ref{fig:Signals' shapes p4581}). The third experiment (p4679) shows aperiodic seismic cycles that are much more challenging to predict.  For each experiment we focus on a data segment of 30 to 50 events shown in the red boxes in Figure \ref{fig:ALLexp}.

\section{Prediction and forecasting models}
We use DNNs by adopting three of the most well-known sequence modelling approaches:  LSTM (Long short-term memory) \citep{ref:Hochreiter1997}, TCN (Temporal Convolutional Network) \citep{ref:Bai2018} and TF (Transformer Network) \citep{ref:Vaswani2017}. 
Figures \ref{fig:architectures_main_prediction} and \ref{fig:architectures_main_forecasting} and Appendix \ref{s:Adopted Deep Neural Network architectures} provide a summary of our models. 

\subsection{Input, output and model performance}
\label{ss:Features, Target and performance}

We use measurements of fault zone shear stress and radiated elastic energy from AE as ML features. Model input consists of statistical measures of the continuous seismic data generated from AE (for the prediction task, see Section \ref{ssec:PART 1- Prediction model}) or the lab measurements of shear stress (for forecasting, see Section \ref{ssec:Forcasting}). The model outputs (different for prediction and forecasting) are compared to the input measurements of ground truth. 
Because we are dealing with time series of physical processes it is inappropriate to shuffle data in time. This limitation holds for both our prediction and forecasting tasks. 

\begin{figure}[H]
\centering
    \includegraphics[width=0.78\textwidth]{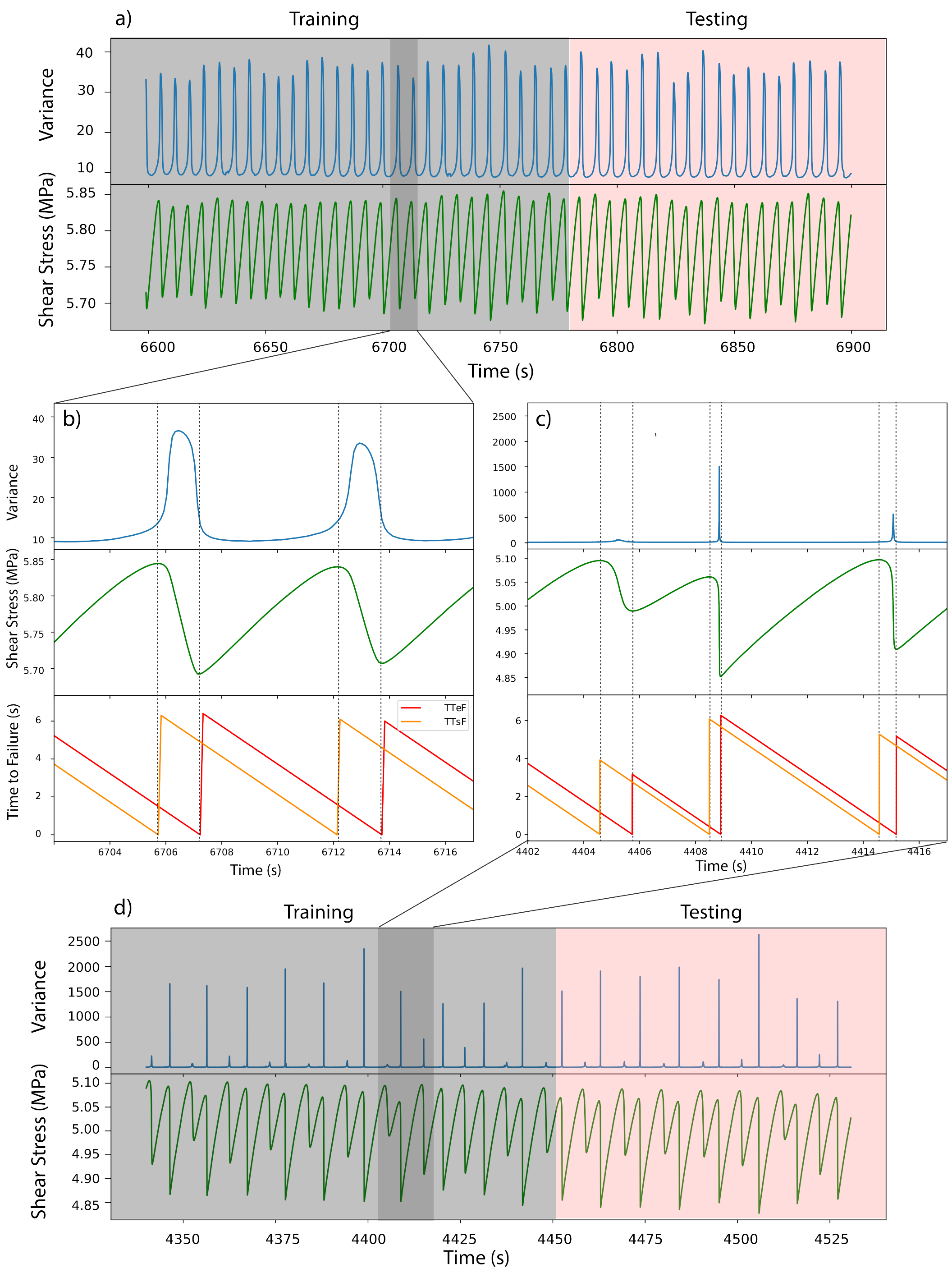}
    \caption{Data for a series of labquakes showing the training and testing phases. (a) Quasi-periodic events during experiment p5198. Panels (b) and (c) are zooms of (a) and (d), respectively, but also show time to the start of failure TTsF (i.e. time of maximum shear stress preceding labquakes) and time to the end of failure TTeF (i.e. time of minimum shear stress after events). (d) Labquakes from experiment p4679 showing complex behavior including period doubling and both fast and slow events. For Panels (a) and (d), variance of the continuous seismic data is plotted above shear stress. Note that we just plot the variance of Channel 1's AE; the variance of Channel 2 is similar.   
    Gray and pink shading show examples of training and testing data split (note that we use a different split for prediction vs. forecasting).
    }
    \label{fig:Signals' shapes}
\end{figure}

As in \cite{ref:RouetLeduc2017}, we consider the variance of the continuous acoustic signal as the most important ML feature for the prediction task. 
An example of raw data for the acoustic signal and shear stress is shown in Figure \ref{fig:pict} while the variance is shown in Figure \ref{fig:Signals' shapes}.

For the ML prediction task, our model predicts as output, often called target, the Shear Stress and/or the Time To Failure (TTF), defined as the time remaining before the next labquake, derived from the shear stress time series. We predict both the time of start of failure (TTsF) and end of failure (TTeF). The former derives from the time of maximum shear stress preceding labquakes (Figure \ref{fig:Signals' shapes}, orange line). That is, TTsF is 0 at the maximum shear stress and it increases backward in time within the lab seismic cyle. TTeF is derived from the minimum shear stress after an event (Figure \ref{fig:Signals' shapes}, red line). TTeF is 0 at the stress minimum and it increases backward in time within the lab seismic cycle.  
We show both TTsF and TTeF along with shear stress and the variance of the continuous AE record for several lab seismic cycles in 
Figure \ref{fig:Signals' shapes}. 
Note here the stepwise nature of TTsF and TTeF as linear functions that our models are designed to predict. 
Note also the differences in the seismic cycles for our experiments: p5198 shows a somewhat periodic sequence of labquakes with lower acoustic energy compared to p4679, which has complex seismic cycles and greater acoustic energy release (Figure \ref{fig:Signals' shapes} and \ref{fig:Signals' shapes p4581}). 


To calculate AE statistics and determine features for our ML algorithms, we use a moving window on the continuous data which were acquired at 4 MHz (Table \ref{tab:win_len_shift}). 
We adjust this window for each experiment based on the seismic cycle duration (Figure \ref{fig:Signals' shapes}), so as to preserve all of the shear stress evolution while not losing small differences between various cycles. In particular, we adopt window lengths similar to those used previously \citeg{ref:RouetLeduc2018, ref:Hulbert2019} so that we can make reliable comparisons of model performance.

In previous work they use a so called "subwindows" procedure to address non-single valued functions. This means that one computes the ML features from the continuous seismic signal in two, slightly offset time windows. This is done because the targets (both shear stress and TTF) are symmetric in time within the lab seismic cycle. By using two subwindows the algorithm is able to differentiate between the loading and the slipping part of the seismic cycle. However, our DL models provide a more robust solution to this problem and one that has much more transportability. We do not need subwindows because the LSTM layer carries information about where we are within the seismic cycle. Thus we just use one window for each AE recording.\\

For our forecasting work we use shear stress measurements, with past data as input and future times as target output. For forecasting, we do not use TTF data but rather forecast shear stress and therefore failure events directly from the shear stress. We smooth the signal in this case too, using a running average window. Then we reduce the resolution to make the length of the sequences suitable for our ML models (details in Table \ref{tab:win_len_shift}). In particular, we sub-sample the signal by a factor of 100 (upon smoothing to limit aliasing). 

A typical ML protocol requires at least two subsets of data, one for training, to learn the model parameters, and one for testing, to measure the generalizability of the results to unseen data (Figures \ref{fig:Signals' shapes} and \ref{fig:Signals' shapes p4581}). We also use a validation dataset to evaluate and optimize the model fit during training and to avoid overfitting. Validation is done with a small portion of data (generally $10\%$)  not used for training nor for testing. The split among the three segments preserves the statistical properties of the dataset. 

A misfit (loss) function is required to build the ML algorithm and optimize the model parameter weights. We use a root mean squared error (RMSE) loss function to compare output prediction and ground truth data. From this comparison, we iteratively adjust model parameters (weights and biases) to minimize the loss. Features and target are generally normalized because their ranges typically vary widely and because normalization helps DL algorithms converge with better results.  The final performance of our ML/DL models is evaluated with respect to ground truth in the form of our lab data (Table \ref{tab:exp results pt1} and \ref{tab:exp results pt2}).  For evaluation metrics we use both the coefficient of determination $R^2$ and the RMSE. 
Summarizing, RMSE is used as loss, following common practise in ML, but also as a test metric representing the discrepancy between model output and ground truth. 

\subsection{Prediction}
\label{ssec:PART 1- Prediction model}

\subsubsection{Problem definition}

Our goal is to predict the present value of shear stress or TTF, as target variables (e.g., $y_t$ where t is the present time) given current  information about AE, as an input variable, (e.g., $x_t$) and it's recent history (e.g. from $x_{t-k}$ to $x_{t-1}$). We note that the temporal evolution of the AE signal during the lab seismic cycle differs for our range of experiments, in particular during the creep phase preceding labquakes (Figures \ref{fig:Signals' shapes} and \ref{fig:Signals' shapes p4581}). 
Previous works have observed that AE statistics (in particular signal amplitude variance and higher-order moments) are highly effective at predicting laboratory earthquakes \citep{ref:RouetLeduc2017}. 
Thus we begin by following this approach. 
However, whereas previous works focused on data from only one acoustic sensor, we use data from two sensors, one on either side of the DDS assembly  (Figure \ref{fig:pict}). This was done by simply using each data stream as a separate input for the continuous seismic signal. Using multiple sensors will clearly be valuable when these techniques are applied to tectonic faults and/or to event location. We did limited testing of the lab data and did not see dramatic improvement over what one would expect by having additional data to constrain the model loss during training.

We consider for the ML process two channels of AE variance $x$ as input, from time $t-k$ to time $t$, and we predict the target $y$ (shear stress or TTF) at time $t$ using the input variable and its recent history. We want to learn the ideal function $f$ that maps our input $x$ into the desired output $y$. With ML we can approximate that function with $\hat{f}$, leveraging input data, and then approximating the variable of interest $\hat{y}.$
In particular, the estimator $\hat{f}(x_{t-k},...,x_{t})$ makes predictions for sample $y_{t}$. 
The quantity $k$ determines the length of the input sequence ($k+1$ in this case) and represents the LSTM's memory in the internal state.
We chose the hyperparameter $k$ using an ad hoc iterative approach (Section  \ref{sss:Best length for the past memory of LSTM}).

\subsubsection{Deep learning model}

We use a DL model based on the implementation of Levinson et al. (\url{https://www.kaggle.com/c/LANL-Earthquake-Prediction/overview} \url{https://www.linkedin.com/pulse/my-team-won-20000-1st-place-kaggles-earthquake-corey-levinson/}). This model, from Team "The Zoo," won the Kaggle competition for lab earthquake prediction \citep{ref:Johnson2021}. We optimize the model for our purposes, including varying hyperparameters and other minor changes, i.e. we adapt the model to allow it to predict one or more targets, as desired. The modified hyperparameters are: 1) using misfit (loss) of 'mean squared error' instead of 'mean absolute error', 2) in model.compile(): we disregard the parameter loss\_weights, that is set to "none", because our model doesn't use a multitarget approach, and 3) in callbacks: monitor='val\_loss', mode='min' instead of monitor='val\_regressor\_mean\_absolute\_error'

\begin{figure}[H]
\centering
    \includegraphics[width=\textwidth]{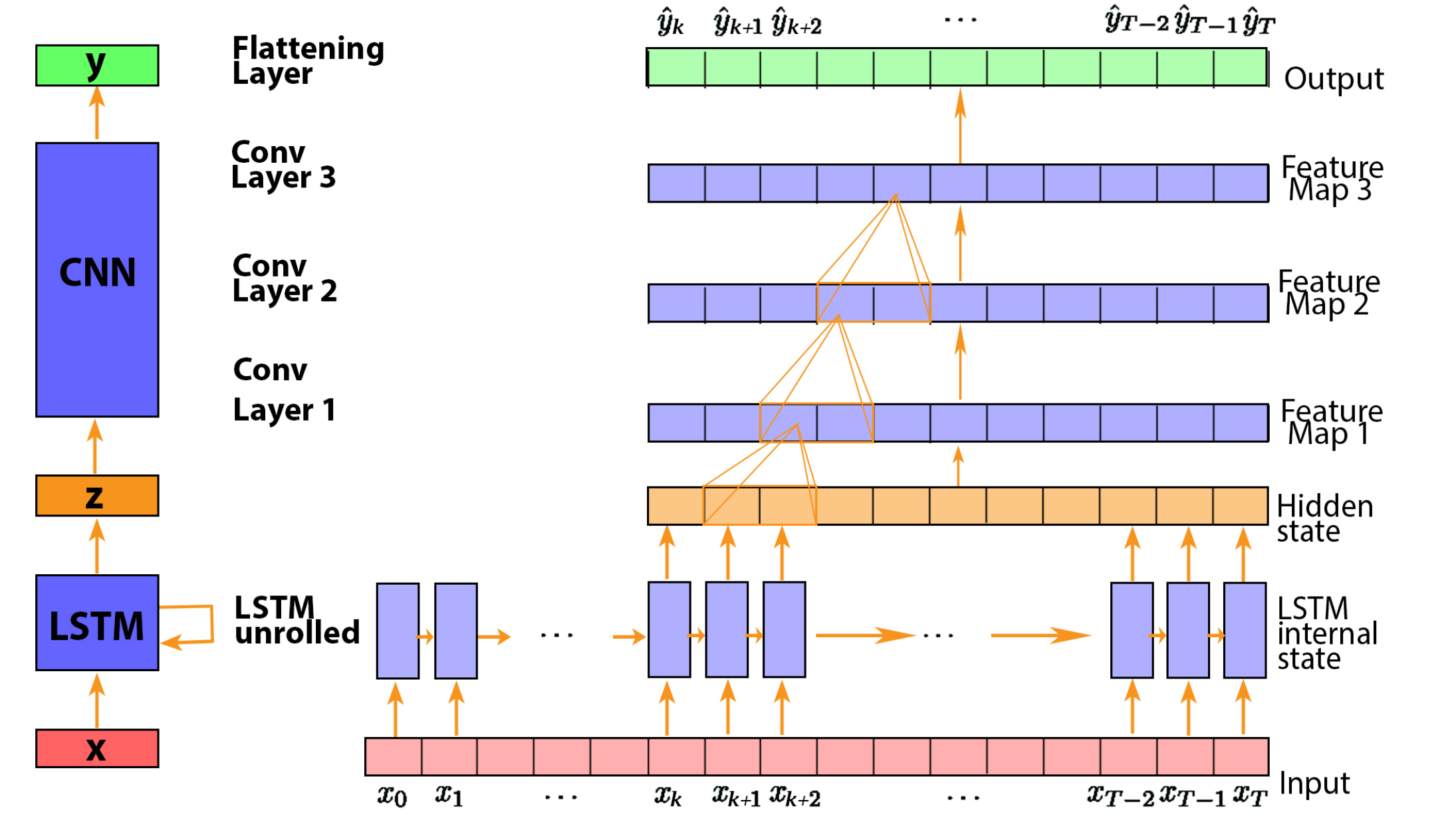}
    \caption{
    Schematic model of our combined LSTM and CNN architectures. LSTM scans the input to produce an embedding (lowest level above input). The LSTM layer is followed by three successive CNN layers (see far left) that make the predictions. The input for each layer is the output of the previous layer.
    The LSTM representation (center and right side) is unrolled in time, which means that each $\hat{z}_t$ (hidden state z at time t) is predicted by considering information coming from the whole sequence from 0 to t. Thus, the LSTM output starts from $k$, which represents the LSTM memory length. Further information, as well as discussion about best $k$ value selection, are in Section \ref{sss:Best length for the past memory of LSTM}. 
    The convolutional layers are used to extract features from the input signal (z in this case). The mathematical operation of convolution is performed between the input signal and a convolutional sliding filter of dimension 2. Here, the output of each layer is called a feature map, which gives information about the signal features.
    Red and pink denote input; green denotes output. Blue denotes hidden states and orange denotes the connections and components inside hidden states. Further details are provided in the Supplementary Section \ref{s:Adopted Deep Neural Network architectures}.}
    \label{fig:architectures_main_prediction}
\end{figure}

As shown in Figure \ref{fig:architectures_main_prediction}, our architecture is a combination of an LSTM layer, which scans the input sequence and produces an embedding, and three stacked convolution (CNN) layers (Figure \ref{fig:architectures_main_prediction}). 
This LSTM + CNN architecture extracts the patterns of the sequence and predicts the target. We predict the target for each time t separately, not as a sequence. Therefore, it is sufficient to specify the number of targets and thus we predict N targets for each time step. The model has a total of 421277 parameters.
We implement the model in Keras (\url{https://keras.io/}), an open-source software library that acts as an interface for the TensorFlow machine learning library (\url{https://www.tensorflow.org/about}). We used Keras to be consistent with the work of Levinson et al.  

\subsection{Forecasting}
\label{ssec:Forcasting}
\subsubsection{Problem definition}
\label{ssec:PART 2- Forcasting autoregressive model}
Autoregressive forecasting methods are distinct from existing predictive models. They predict not only the current state of the target (e.g. $y_t$, where $t$ is the present time), but also future steps (e.g. from $y_{t+1}$ to $y_{t+N}$, where $N$ is the length of the sequence to predict) given past information (e.g. from $x_{t-k}$ to $x_{t-1}$). 
In particular, in the forecasting problem the input and output variables need to be of the same nature; that is,  
we can input  shear stress data from time $t-k$ to $t-1$ and predict shear stress in the future from time $t$ to $t+N$ . 
This type of DL model uses an autoregressive technique (AR) in which regression-based training occurs on the input variable itself.
Autoregressive models predict their own input features one step at a time and keep predicting longer-term future horizons by using the previous output as input for the next step (Figure \ref{fig:architectures_main_forecasting}). 
The estimator $\hat{f}(x_{t-k},...,x_{t-1})$ makes predictions for samples $\hat{x}_t$ from measurements of the target $x_{t-k},...,x_{t-1}$ and continues predicting autoregressively $\hat{x}_{t+1}, \hat{x}_{t+2}, ..., \hat{x}_{t+N}$, using the previous outputs $\hat{x}_{t}, \hat{x}_{t+1}, ..., \hat{x}_{t+N-1}$.

This proceeds iteratively as:
\begin{flalign*}
&STEP\ 0: \hat{f}(x_{t-k},...,x_{t-1})=\hat{x_{t}} \\
&STEP\ 1: \hat{f}(x_{t-k+1},...,x_{t-1},\hat{x_{t}})=\hat{x}_{t+1}\\
&...\\
&STEP\ i: \hat{f}(x_{t-k+i},...,x_{t-1},\hat{x_{t}},...,\hat{x}_{t-1+i})=\hat{x}_{t+i}
\end{flalign*}
where $i=0, ..., N$.
During AR training we apply the Teacher forcing technique. Teacher forcing is known to introduce a so-called exposure bias \citep{ref:He2019}, since the training procedure focuses on predicting the next step, while at inference the model predicts from history and from its own (auto-regressive) predictions. However, it has been shown that the exposure bias may effectively be neglected in most cases \citep{ref:He2019}.

With our use of teacher forcing, the prediction at time $t+1$ uses ground truth at $t$ rather than the prediction of $t$, and proceeds iteratively as: 
\begin{flalign*}
&STEP\ 0: \hat{f}(x_{t-k},...,x_{t-1})=\hat{x_{t}} \\
&STEP\ 1: \hat{f}(x_{t-k+1},...,x_{t-1},x_{t})=\hat{x}_{t+1}\\
&...\\
&STEP\ i: \hat{f}(x_{t-k+i},...,x_{t-1},x_{t},...,x_{t-1+i})=\hat{x}_{t+i}
\end{flalign*}

Thus, we introduce and accept differences between the train and test protocol (i.e., applying the "teacher-forcing" during training) 
but we still leverage the full data set. 
This training is more efficient 
because data are more informative than the predicted values.
Model validation and testing are done without teaching-forcing, because we lack ground truth in those cases.
The AR approach has several advantages. First, it is self-supervised and does not require labelling. This also extends to TTF: in fact, TTF is computed from the actual stress time series. So it is computed from the actual sequence of stress values for the (self-supervised) training and it is computed from the forecast stress values at test. 
Since TTF is a quantity that is manually built, this may be prone to ambiguity due to pre-processing.
Also, AR has the potential to predict beyond the TTF estimates, by  predicting multiple cycles into the future. 

For AR training we set the time interval for input history $k$ (from $x_{t-k}$ to $x_{t-1}$) as the "steps in" variable. Then to establish the prediction time 
(from $y_{t}$ to $y_{t+N}$) we use length $N$ as the "steps out" variable.
The values of $k$ and $N$ for training are also used for validation and testing. While it is possible to increase 
N and extend the forecast horizon, we expect a deterioration in performance for values greater than the $N$ used for training, and we explore such work below.
We implemented the AR work in Pytorch, an open source machine learning library based on the Torch library (\url{https://pytorch.org/}). Additional details of the procedure are provided in the Supplement.



\subsubsection{Forecasting with LSTM}
\label{sssec:PART 2- AR LSTM}
A key component of an LSTM (Figure \ref{fig:unrolled RNN_LSTM_remake}) is the memory cell that regulates information flow using three gates (i.e, Forget gate, Input gate, Output gate). The Forget gate deletes useless information by not releasing it to the next stage. The Input gate regulates new information into the network. The Output gate decides which part of the memory to output from the cell. During the training process, inputs and model weights pass to all gates, which in turn are connected to the self-recurrent memory cell (Figure \ref{fig:unrolled RNN_LSTM_remake}).
Our model has 3 stacked layers with size 300, for a total of 1808701 parameters (see Supplement \ref{ss:Long Short Term Memory (LSTM)} for details).

\subsubsection{Forecasting with Temporal Convolutional Networks (TCN)}
\label{sssec:PART 2- AR TCN}

Temporal Convolutional Networks (TCN) 
(Figure \ref{fig:tcn from Bai_a_b_remake}) are 
Convolutional Neural Networks (CNN) that have been adopted for sequence modeling because of their performance \citep{ref:Dessi2019}. 
TCN consists of causal and dilated 1D convolutional layers with the same input and output lengths 
\citep{ref:Bai2018}. Here, the term causal refers to the fact that convolution is applied 
only with the present and past elements but not the future. The term dilation in the context of a convolutional layer refers to the distance between the elements of the input sequence that are used to compute one entry of the output sequence: this increases the receptive field in each layer
, making it possible to model long temporal patterns. 
In sequence modelling, TCN can be viewed as the process of sliding a 1D-window over the sequence to predict the part of the sequence with the length of the receptive field, using the chosen kernel. Such predictions are passed to the subsequent layers and the procedure continues until the receptive field has the size of the input sequence.
The convolutions in TCN can be parallelized at training, because the filter that extracts the features, used in each layer, can be computed jointly. 
RNNs instead need to be unrolled during backpropagation and can not be parallelized. 

We adopt a model composed of three convolutional 1D layers and scan the sequence in a causal fashion, with dilation=1.
Our models have a hidden size of 64 in the first layer and 256 in the second layer. The last layer has the dimension of the output, which is 1 in this case (because the model forecasts the next step). The model has a total of 29761 parameters, which is 2 orders of magnitude less than what we use for LSTM.

\subsubsection{Forecasting with Transformer Network}
\label{sssec:PART 2- AR Transformer Network}

Our Transformer Network (TF) consists of a modular architecture with an encoder and a decoder that contains three blocks: attention modules (self-attention and encoder-decoder attention), a feed-forward fully-connected module, and residual connections (Figure \ref{fig:transformer model Vaswani_remake}). The network's capability to capture sequence non-linearities lies mainly in the attention module. TF maintains the encoding output (memory) separate from the decoded
sequence, which means that training is parallelizable. TF is parallelizable also because of the absence of the internal status (as in LSTMs), so interactions between input and output are direct and do not need recursion for loops of backpropagation.

The model we adopt is based on the work of \cite{ref:Giuliari2020}. 
This model has  $d_{model}$=128, 2 layers and 4 attention heads, for a total of 663938 parameters. We use RMSE and train the network via backpropagation with the NoamOpt optimizer; dropout value of 0.1 (Supplementary  Material \ref{ss:Transformer network}).

Although TF is often favored for its high performance, a weakness is the need for huge training datasets. 
This explains the poor performance of TF for our experiments. 
Thus to improve TF we pre-trained using a sine function as input. This allows the model to learn the oscillatory behavior of the experiments. We used a pre-training dataset of 1e7 rows to describe a long set of sine waves with the same resolution of our lab data. The sine function roughly matches our lab data with amplitude equal to one frequency of 0.1 Hz and sample rate of 1000 Hz. This approach shows that 1 or 2 epochs are enough (an epoch is one complete pass of the training dataset through the algorithm) to pre-train (Section 9.1.4). After pre-training, we fine-tune the model using training data
for each experiment.

\section{Results}

\subsection{Prediction}
\label{ssec:part 1 results}

Shear stress and time to failure  are reasonably well predicted by the LSTM+CNN architecture (Figure \ref{fig:architectures_main_prediction}). 
The predictions match ground truth with an accuracy > $93 \%$ (Table \ref{tab:exp results pt1}). 
Figure \ref{fig:LSTM performance} shows model predictions and indicates that  
long term memory length is a key hyperparameter. 

\subsubsection{Best length for the past memory of LSTM}
\label{sss:Best length for the past memory of LSTM}
We investigate LSTM long term memory duration $k$ using both RMSE and $R^2$ to assess performance. The optimum length for $k$ is about one seismic cycle (Figure \ref{fig:steps performances}).
Note that the the red lines show the optimum $k$. 
For experiments p4581 and p5198 we adopt a lower resolution because of event periodicity and because we compute variance every 0.1 seconds 
(see Section \ref{ss:Features, Target and performance} and Table \ref{tab:win_len_shift}). Thus, one seismic cycle corresponds to about 100 data points and the best length for the observation is $k=70$ or  $7$ seconds. For p4679 we adopt a higher resolution, 0.003 s, to describe both fast and slow events. Here, one seismic cycle corresponds to about 1700 data points, and the optimum value of $k=2000$  ($\sim6$ seconds) for predicting  shear stress (Figure \ref{fig:steps performances}). For predicting TTF, the optimum value of $k=1100$ ($\sim3$ s). 

We found high $R^2$ values for all three experiments using an observation length of about one seismic cycle. This suggests that there is a saturation of performance at one seismic cycle, so this may be sufficient to understand the signal. After one cycle there is a decrease in performance, due to experiment aperiodicity and LSTM memory limitations.

LSTMs struggle to learn long-term trends. The presence of slow and fast events in p4679 requires higher resolution to calculate AE variance.   Compared to sequences of quasiperiodic events, p4679 requires about a factor of 10 more data. 
Long seismic cycles with rapid stress drops represent challenging conditions for LSTM. The problem arises because of observation lengths and the fact that Forget gates tend to remove too much information. 

\subsubsection{Prediction dataset split}
\label{sss:Prediction dataset split}
We train and validate with $70 \%$ of the data and test with $30 \%$  (Figure \ref{fig:Signals' shapes}). Validation data were chosen randomly from the first $70 \%$ of the data and this value ($10 \%$) was removed from the training set. Thus the final division was: $63 \%$ for training, $7 \%$ for validation and $30 \%$ for testing. Details in Table \ref{tab:Training validation testing split}.
We normalized our data using: $X_{norm} = \frac{X-min(X_{train})}{max(X_{train})-min(X_{train})}$. 

\subsubsection{Prediction results}
\label{sss:prediction_results}
Figure \ref{fig:LSTM performance} and Table \ref{tab:exp results pt1} summarize prediction results. Black lines show measurements and the colored lines (green, yellow and red) are predictions for shear stress, TTsF and TTeF, respectively. Note that the predictions are quite accurate. 
The model is able to accurately predict shear stress,  with $R^2$ > 0.9. Also TTF predictions are accurate ( $R^2$ up to 90 \%), even if noisy in a few cases. 
We observe a general trend of better performance for TTeF than TTsF. Also the performance is better for experiments p5198 and p4679 than for p4581. Figure \ref{fig:Signals' shapes} shows that TTeF is maximum where shear stress is minimum and variance is maximum. The peak variance scales directly with stress drop amplitude, with smaller values of peak variance preceding labquakes with smaller stress drop. 
As a consequence of smaller stress drop the time to reach a critical failure stress is smaller for constant loading rate,. Indeed, the maximum shear stress and the slope of the restrengthening phase are quite 
similar in all the seismic cycles. 
In contrast, the values of TTsF derived from maximum shear stress show greater variability --possibly because of creep prior to the stress drop. 

We used only AE variance as model input to predict shear stress and TTF and 
found the same result for data from each AE sensor. Our model performance for each experiment exceeds the state-of-the-art (Table \ref{tab:exp results pt1_VS_XGBoost}) as defined by existing works \citep{ref:RouetLeduc2017,ref:RouetLeduc2018}.
While this suggests that we could further improve model performance, by using more complex statistical features, rather than just variance. However, we did not pursue this direction and instead focused on new DL models. 




\subsection{Experiments on Forecasting}
\label{ssec:part 2 results}
Autoregressive forecasting models provide a method to predict shear stress variations based on past observations.
We test AR models using our three DNN architectures: LSTM, TCN and TF. 
We analyzed all experiments but focus here on p4679 because of its complex behavior and aperiodic seismic cylces. For this task, we decimated shear stress data to $dt=0.1$s and we apply an average running mean (Table \ref{tab:win_len_shift}). 
Although this reduces the original data resolution, 
downsampling is necessary for computational load and to limit the size of the step-in hyperparameter, which is particularly important for representing multiple seismic cycles.
To establish the AR models we use $k=200$ (20 seconds) which is a value that both describes seismic cycles and fits GPU memory limitations.

Our data include many seismic cycles but this number is actually quite small for training DL models. Thus,  we forecast using overlapping windows. 
To compare performance of our DL architectures (LSTM, TCN and TF) for each experiment, we set up the model to predict 100 data points (10 seconds), which corresponds to 1-2 seismic cycles depending on the experiment. 
See Figure \ref{fig:pt2 AR performance} for a sketch of these overlapped sequences. 
Shear stress forecasting is possible, although prediction accuracy varied between the DL models.

\subsubsection{Forecasting dataset split}
\label{sss:Forecasting dataset split}
Our AR work used the same normalization as for prediction model, described above, while the division into train / validation/ test subsets is different. The training set consists of the first 70$\%$ of the data and the testing set consists of the last 20$\%$. Model validation was done with the remaining $10 \%$ of the data. Details in Table \ref{tab:Training validation testing split}. Model performance was evaluated for future predictions in each window (Figure \ref{fig:pt2 AR performance}). 
We also do an average among windows to measure overall performance (Table \ref{tab:exp results pt2}). 
Different data segments result in different performances, thus
in Figure \ref{fig:pt2 AR performance} we show predictions for a range of different starting locations within the data stream. Note that  AR models perform well for a range of starting times and data segments.

\subsubsection{LSTM forecasting results}
\label{sssec:part 2 LSTM results}

LSTM models generally produced the poorest predictions of future stress states (Figure 7). Our AR forecast results are reasonable for only p5198, which has the least complex seismic cycles. For the other experiments LSTM did not provide accurate predictions. 
While LSTM is generally good for time series forecasting, because of its flexibility and optimization, it has trouble with data from our experiments because of the data density, even if significantly decimated (dt=0.1 s), and seismic cycle length (of order 10 s). 
LSTM works well with fewer data points. For example, in the work of \cite{ref:Giuliari2020} they have 8 data points in input and 12 data points in output, whereas we have 200 in input and 100 in output. In essence, for our experiments the LSTM network "forgets" the past too quickly and does not predict accurately the signal in the future. 

\subsubsection{Temporal Convolution Network forecasting results}
\label{sssec:part 2 TCN results}
While TCN is the simplest of our models, it produced some of the best AR results. The TCN models provide the best compromise between training complexity and the number of parameters required. TCN requires only one tenth of the parameters needed for LSTM and TF.
From Table \ref{tab:exp results pt2} and Figure \ref{fig:pt2 AR performance} we can appreciate the TCN capability for forecasting. Results for experiment p5198 are quite good as are those for the more complex case of p4679. 
For p4581, TCN is the best of the tested models based on average forecast accuracy. 
However this experiment turned out to be challenging for each of the AR models as none of the forecasts were very good (Figure \ref{fig:pt2 AR performance}), possibly due to the lack of appreciable preseismic creep in this experiment. 

\subsubsection{Transformer Network forecasting results}
\label{sssec:part 2 Transformer Network results}
TF is the second best model, after TCN. It is not the best on average in any experiment, however with the flexible attention focusing it captures higher order variations of the data. For the same reason it does not forecast well the average behaviour of the signal, as a simpler model like TCN does. 
TF is the most complex among the tested models for optimization and training, and it is also the most flexible, because of its complex connections. It requires a lot of data
to start working, and for this reason we designed a pretraining stage in which we fed the model sine waves with a frequency similar to our data, as described in \ref{sssec:PART 2- AR Transformer Network}.

In several segments of p4581 TF is capable of detecting irregularities in the seismic cycle (Figure \ref{fig:pt2 AR performance}).
However, in other places TF tries to predict irregularities that are noise rather than signal. One example of this occurs close to the local maxima in Figure \ref{fig:pt2 AR performance} for p4581. Noise features like this caused trouble for TF. On the other hand, TF did well 
to forecast the aperiodicity of p4679 (Figure \ref{fig:pt2 AR performance}). For p5198, TF 
did well, possibly because the behavior is quite periodic.
This is an indication that data overfitting during training may be a problem. We note that the improvement based on simple pre-training using a periodic signal provides some interesting future direction.

\subsubsection{Forecasting results: model  comparison and analysis}
\label{sssec:Forecasting results: models comparison and analysis}

A key comparison between our AR models is how they perform with respect to the longer term temporal variations of the seismic cycle.  Thus we evaluate results for several 
different data segments within the test set (Figure \ref{fig:pt2 AR stress and metrics}). 
Note that the performances in Table \ref{tab:exp results pt2} are averaged for the entire test set. In Figure \ref{fig:pt2 AR stress and metrics} we plot results for windows starting at several positions within the data.
We can see that all the models have high variability depending on the tested window, hence the capability of the model in predicting the signal depends on the shape and level of irregularity of the input-output window. 
As expected from our summary of general results, LSTM produced the worst AR forecasts. The other three models are in general able to forecast the signal with quite low variability. 
We note also that for p4581 the performance decreases with time in some areas simply because the signal becomes more challenging, for example where there are irregularities before peak stresses as seen on the left in Figure \ref{fig:pt2 AR stress and metrics}.

An important question for the AR forecasting is that of how far into the future these models can predict and in particular if they can predict beyond a single seismic cycle. 
Figure \ref{fig:pt2 performance_degradation_avg} summarizes such testing. Here we plot the average of all the performance values for each forecast time in the future. Our models are trained to forecast 100 data points in the future (white section), however we extended these to forecast 200 data points in the future (gray section). LSTM simply gets worse as time is extended. TCN decreases in performance slowly,
whereas TF has a peculiar behaviour.
For forecasting times from 0 to 100, TF is similar to TCN, while afterwards it becomes suddenly worse. This is because in the training phase TF forecasts from 0 to 100, but the behavior is somewhat different in the next seismic cycle -- from $[101, 200]$. 
This is perhaps surprising given that TF is renowned for generalization. Here, it is possibly the result of data complexity. This is clearly an important question for future work.


\section{Discussion}

The first part of our work was devoted to evaluating DL models based on a combination of LSTM and CNN. Compared to the state-of-the-art based on existing works we find that DL models perform well. 
As suggested by previous works, we used the variance of the continuous acoustic emissions emanating from the laboratory faults. 
The DL models work reasonably well for all tested experiments, however results are better when creep occurs before the mainshock (i.e. as in p5198 and p4679). This is perhaps not surprising given that creep represents fault slip which could result in micro fracture and breakage of frictional asperity contacts.  Moreover, the shear stress curves reflect creep and AE via the gradual reduction in the rate of increase prior to reaching a 
peak at the onset of a labquake (Figure \ref{fig:Signals' shapes}). 
In the absence of creep, such as in p4581 (Figure \ref{fig:Signals' shapes p4581}), the acoustic variance is initially very low and increases suddenly at the onset of failure. As a result the DL models struggle to identify the exact time of the drop in shear stress. 
This appears to be part of the reason why Time To Start of Failure (TTsF) is always more challenging to predict than Time To End of Failure (TTeF), because the former is zero when the variance is beginning to grow, while the latter is zero right after the variance has first increased and then decreased.  

The limitation of TTF is that it represents the time remaining for just one event. Moreover this quantity is not recorded directly during the experiment in the laboratory. Therefore we manually labelled a suitable dataset for it, before the model training phase.
The autoregressive models solve this problem nicely because they are designed to forecast. 

Our autoregressive forecasting procedure allows one to predict future values of shear stress during the lab seismic cycle. 
To the best of our knowledge, this is the first time such an AR approach has been used for lab earthquake data. 
Here, the innovation is that we can indefinitely forecast into the future without the need for labelling because we use the the same feature (shear stress) for both model input and output. 

We tested several DL networks for the AR forecasting. 
All of the models work at some level. The LSTM network produced the poorest results. In time series forecasting, LSTM is one of the most common models, because of its flexibility, but in our application it has memory problems due to the considerable length of the sequences.
TCN is the simplest model in terms of number of parameters and structure, and it performed better, perhaps because the target (shear stress) during the lab seismic cycle is somewhat periodic. 
TF is also flexible but it is the most complex for optimization and requires a lot of data 
to initiate a reasonable model because of the huge number of parameters to be trained. We dealt with that problem by pretraining the TF models with sine waves, however that represents another step in the processing. 

Of the experiments we evaluated, p5198 was fit best by the DL models. 
Experiment p4679 was challenging because of its aperiodicity and the presence of slow and fast events. However our TCN and TF models were able to  forecast reasonably well. 
The forecasts for p4581 were more challenging, perhaps because of the lack of appreciable fault creep and/or because of noise in the data. Further work is needed to resolve these issues.\\
Overall, DL is capable of understanding the behaviour of the signal and this opens new perspectives in seismology research with the use of AR methods. First of all, this confirms that there are some patterns in the signal that the model is able to recognize in most of the cases. Moreover, the model after training, just needs a limited knowledge about the past of a specific window to forecast reasonably the future. This provides insight on the physical processes and their deterministic nature. 

One good way to advance our work would be to apply the method on real fault data. This would require measurements of shear stress, which are not available. The idea could be to use seismograms directly prior to or during an earthquake and infer the shear stress using the prediction method, such as we did with AE in the lab. This would allow forecasts of the shear stress for future time steps, using the forecasting method. This procedure has the limitation that labquakes are simplistic compared to earthquakes, and this is one of the reasons why it is important to improve our understanding of ML methods for prediction and forecasting.
Another approach would be to simply predict the temporal evolution of the seismogram based on the initial part of the signal. A key question here is then, are lab AE signals uniquely relatable to fault shear zone stress and if so, would that transfer to tectonic faults? Yet another approach would be transfer learning from lab earthquakes to real fault data. 


\section{Conclusions}

We used Deep Neural Networks to predict and forecast laboratory earthquakes and lab measurements of fault zone shear stress based on seismic signals emanating from lab faults. Previous work showed that using the variance of lab seismic signals (from fault zone acoustic emissions) it is possible to predict fault shear stress and the time to failure. We systematically tested a range of DL models with a variety of lab faults and found that our models significantly outperform the state-of-the-art.
Moreover we proved that it is possible to forecast future fault zone shear stress based on the previous history of stress values. 
This result has significant potential because shear stress is representative of the state of the fault and forecasting it in time means implicitly to forecast the timing of the future failure states. To our knowledge, this is the first application of a forecasting procedure with the goal of inferring autoregressively the future shear stress knowing the stress itself in the past. 



\subsection*{Data and resources} 
All simulation input files and the Jupyter notebooks are accessible at GitHub: \url{https://github.com/lauralaurenti/DNN-earthquake-prediction-forecasting}.

\subsection*{Acknowledgments}
We would like to thank M. Denolle and two anonymous reviewers for helpful comments that improved the manuscript. We would also like to thank M. Scuderi, C. Collettini, P. Johnson, and G. Paoletti for helpful discussions. L.L., E.T., and C.M. were supported by European Research Council Advance grant 835012 (TECTONIC). C.M. also acknowledges support from US Department of Energy grants DE-SC0020512 and DE-EE0008763.

\subsection*{CRediT authorship contribution statement}
Laura Laurenti: Conceptualization, Data curation, Formal analysis, Investigation, Writing – original draft, Writing – review \& editing.
Elisa Tinti: Conceptualization, Investigation, Writing – original draft, Writing – review \& editing.
Fabio Galasso: Methods, Data curation, Validation, Supervision, Writing – review \& editing
Luca Franco: Methods, Validation, Writing – review \& editing.
Chris Marone: Conceptualization, Data collection, Investigation, Methods, Supervision, Resources, Validation, Software, Writing – review \& editing.
All authors approve on the submitted article.

\newpage

\bibliographystyle{model2-names}
\biboptions{authoryear}
\bibliography{ref}


\clearpage

\section{Tables}

\begin{table}[ht]
\begin{center}
\begin{tabular}{|c|c|c|c|c|}
    \hline
    \multirow{4}{*}{Target} & GOF  & p4581 & p5198 & p4679 \\
    &    & Glass & Quartz & Quartz \\
    &    & beads & powder &  powder\\
    \hline
    \multirow{3}{*}{Shear Stress} & $R^2$ & 0.9254 & 0.9884 & 0.9574 \\
    & $RMSE$ & 0.0670 & 0.0305  & 0.0519 \\
    \hline
    \multirow{3}{*}{Time To Start Of Failures} & $R^2$ & 0.6317 & 0.9313  & 0.8229 \\
    & $RMSE$ & 0.15844 & 0.0728 & 0.1087 \\
    \hline
    \multirow{3}{*}{Time To End Of Failures} & $R^2$ & 0.8721 & 0.9697 & 0.9200  \\
    & $RMSE$ & 0.0972 & 0.0476 & 0.0723 \\
    \hline
\end{tabular}
\end{center}
\caption{Results for DL prediction of using the continuous lab seismic signal. For each target we show goodness of fit (GOF) in terms of the coefficient of determination $R^2$ and the root mean square error RSME. Shear stress is reasonably well predicted for each experiment. Of the three targets, TTsF is the hardest to predict. Experiment p4581 is the hardest one to predict.}
\label{tab:exp results pt1}
\end{table}

\begin{table}[ht]
\begin{center}
\begin{tabular}{|c|c|c|c|c|}
    \hline
    \multirow{4}{*}{Model} & GOF & p4581 & p5198 & p4679\\
    &    & Glass & Quartz & Quartz \\
    &    & beads & powder &  powder\\
    \hline
    \multirow{3}{*}{TCN} & $R^2$ & 0.3935 & 0.9419 & 0.8273 \\
    & $RMSE$  & 0.1245 & 0.0549 & 0.0732\\
    \hline
    \multirow{3}{*}{LSTM} & $R^2$ & $-$4.5193 & 0.8021 & $-$0.2704 \\
    & $RMSE$ & 0.1521 & 0.0904 & 0.1634 \\
    \hline
    \multirow{3}{*}{TF} & $R^2$ & 0.1172 & 0.8914 & 0.7940 \\
    & $RMSE$ & 0.1460 & 0.0707 & 0.0738\\
    \hline
\end{tabular}
\end{center}
\caption{Results for autoregressive forecasting of fault zone shear stress showing a comparison for each experiment and three models. The goodness of fit (GOF) values are averages computed for all the segments in the testing data. The GOF values vary among the segments as described in the text. }
\label{tab:exp results pt2}
\end{table}



\clearpage

\section{Figures}

\begin{figure}[H]
    \centering
    \begin{subfigure}[b]{\textwidth}
        \centering
        \includegraphics[width=0.8\linewidth]{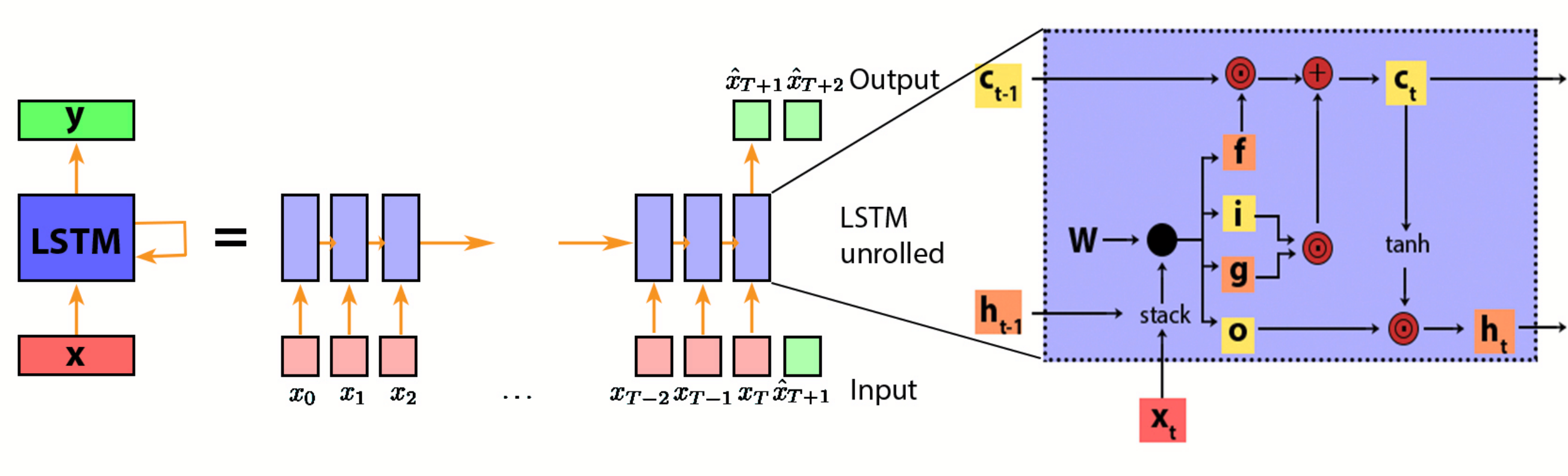}
        \caption{Time-unrolled LSTM \citep{ref:Hochreiter1997}. Zoom at right shows details of the LSTM cell representation. The state-vector ($h_t$) is recursively updated at each t, including the new information coming from the current input and also the cell state ($c_t$).}
        \label{fig:unrolled RNN_LSTM_remake}
    \end{subfigure}
    \begin{subfigure}[b]{\textwidth}
        \centering
        \includegraphics[width=0.90\linewidth]{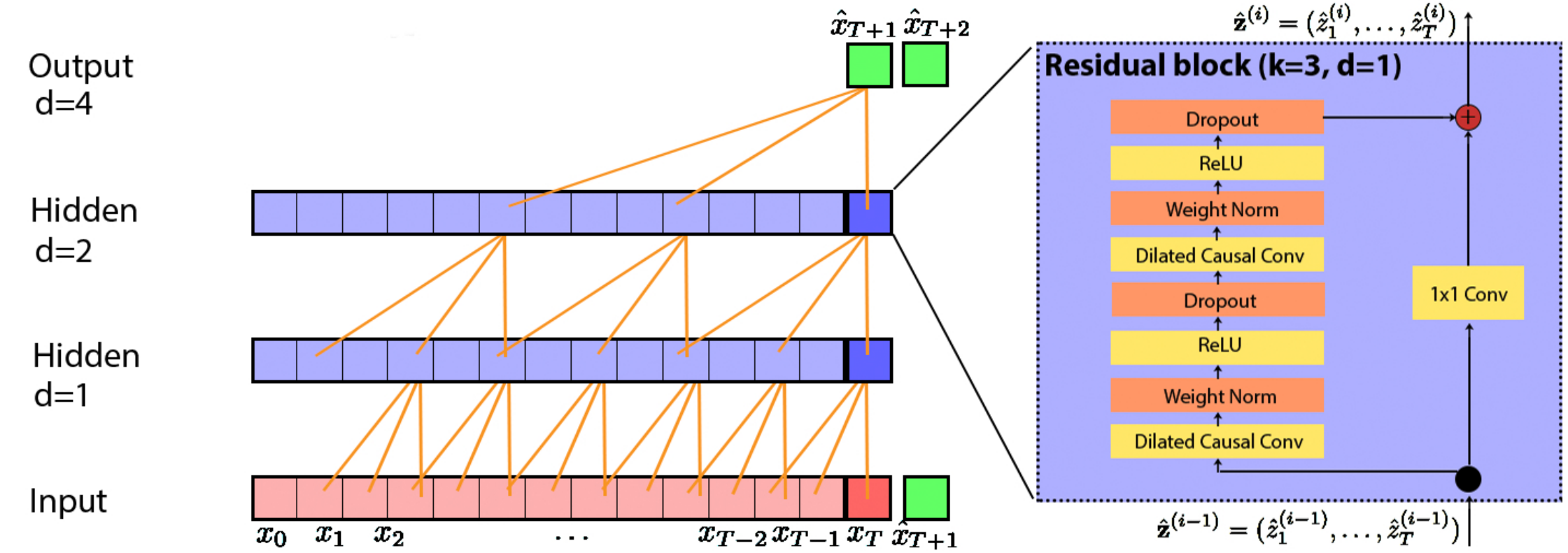}
        \caption{Architectural elements in a TCN \cite{ref:Bai2018}. A dilated causal convolution with dilation factors d = 1, 2, 4 and filter size k = 3. The receptive field is able to cover all values from the input sequence. Zoom on the right shows the TCN residual block. Here a 1x1 convolution is added when residual input and output have different dimensions.} 
        \label{fig:tcn from Bai_a_b_remake}
    \end{subfigure}
    \begin{subfigure}[b]{\textwidth}
        \centering
        \includegraphics[width=0.35\textwidth]{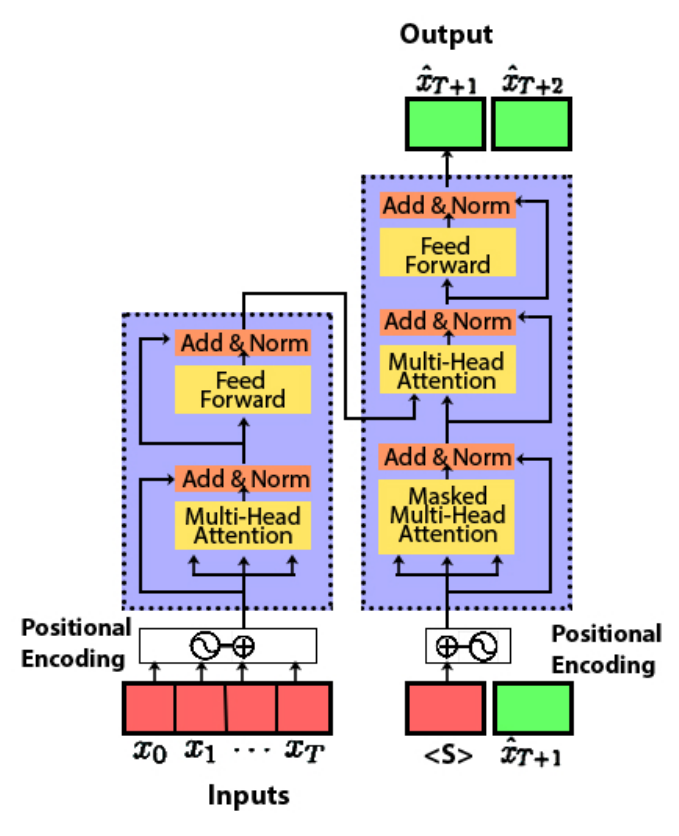}
        \caption{Transformer architecture \cite{ref:Vaswani2017}.}
        \label{fig:transformer model Vaswani_remake}
    \end{subfigure}
    \caption{Schematic showing how TCN, LSTM and TF are used for our forecasting models. In each case, red denotes input, green denotes output, blue denotes hidden states, and orange denotes the connections and components within hidden states. These models are autoregressive (AR). Thus, the inputs are prediction $\hat{x}_{T+1}$, coming from the previous iteration. Further details are provided in the Supplementary Section \ref{s:Adopted Deep Neural Network architectures}}
    \label{fig:architectures_main_forecasting}
\end{figure}

\begin{figure}[H]
    \centering
    \includegraphics[width=\textwidth]{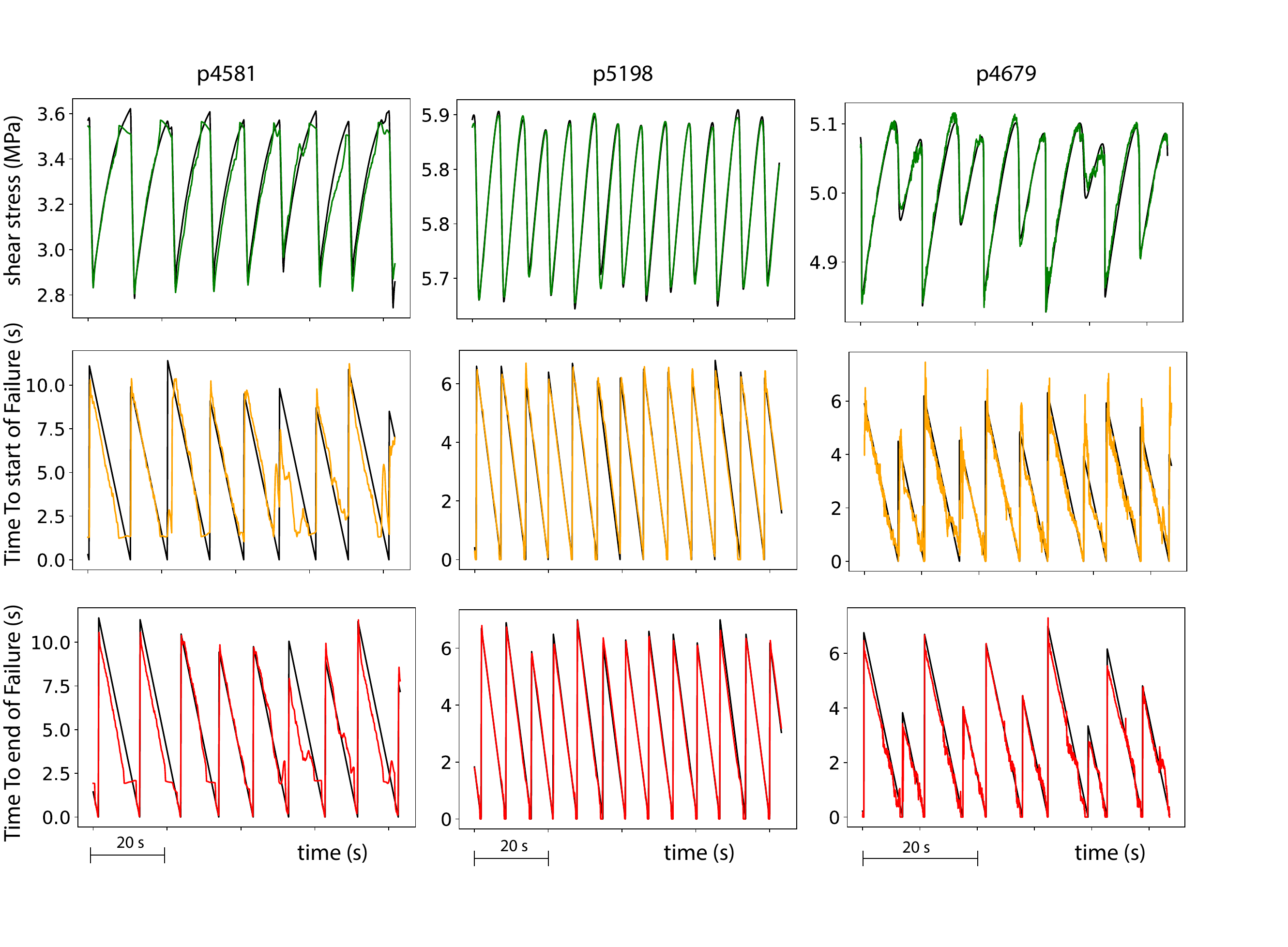}
    \caption{Results of the model predictions for three experiments.
    Black lines show lab measurements (ground truth) and colored lines are ML predictions. Note that shear stress (green lines) and time to end of failure (red lines) are well predicted in all the experiments (see also Table \ref{tab:exp results pt1}).
    Predictions of time to start of failure (orange lines) are also quite good in general, excepting a few sections of p4581. For p4679 we can see that, even if the prediction seems a bit noisy, the behaviour of the function is always well predicted.
    }
    \label{fig:LSTM performance}
\end{figure}


\begin{figure}[H]
    \centering
    \includegraphics[width=\textwidth]{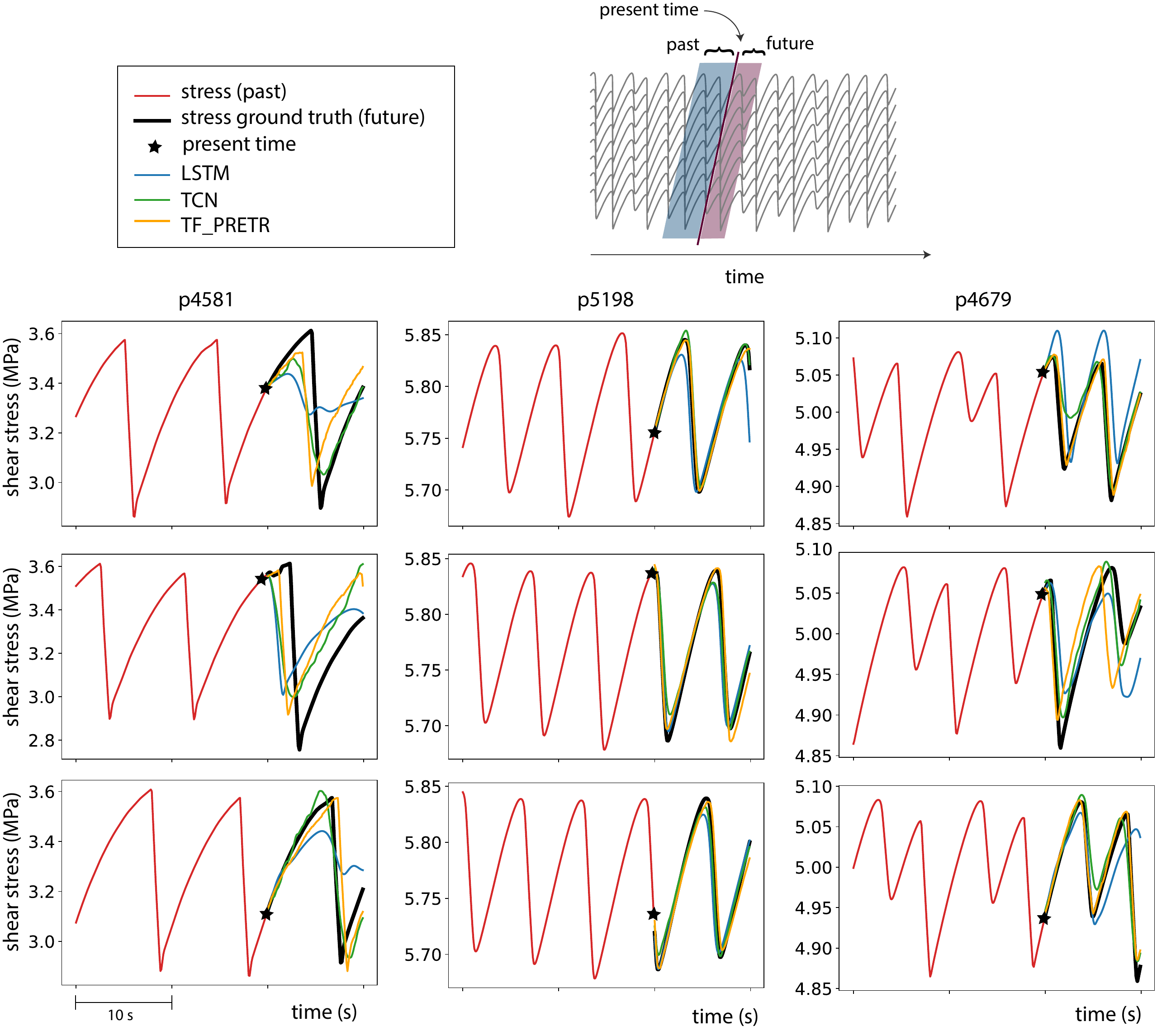}
    \caption{Results of the autoregressive (AR) forecasting models for three experiments. Forecasts vary throughout the experiment, so for each experiment we show three time windows for the forecast. In each case red lines show stress measurements used as model input, black lines show ground truth future stress and colored lines show model output forecasts.  Forecasts are poor for experiment p4581 and the
    accuracy varies significantly from one window to another. For p4581, TF does better than LSTM and TCN at times. 
    The forecasts are best for p5198, where all models predict the target quite well. 
    Experiment p4679 shows a complex set of large and small events and is the most challenging. Forecasts here are quite variable depending on the time interval. However the models are able to predict reasonably well, especially TCN and TF. In each case, LSTM provides the poorest fits. }
    \label{fig:pt2 AR performance}
\end{figure}

\begin{figure}[H]
    \centering
    \includegraphics[width=\textwidth]{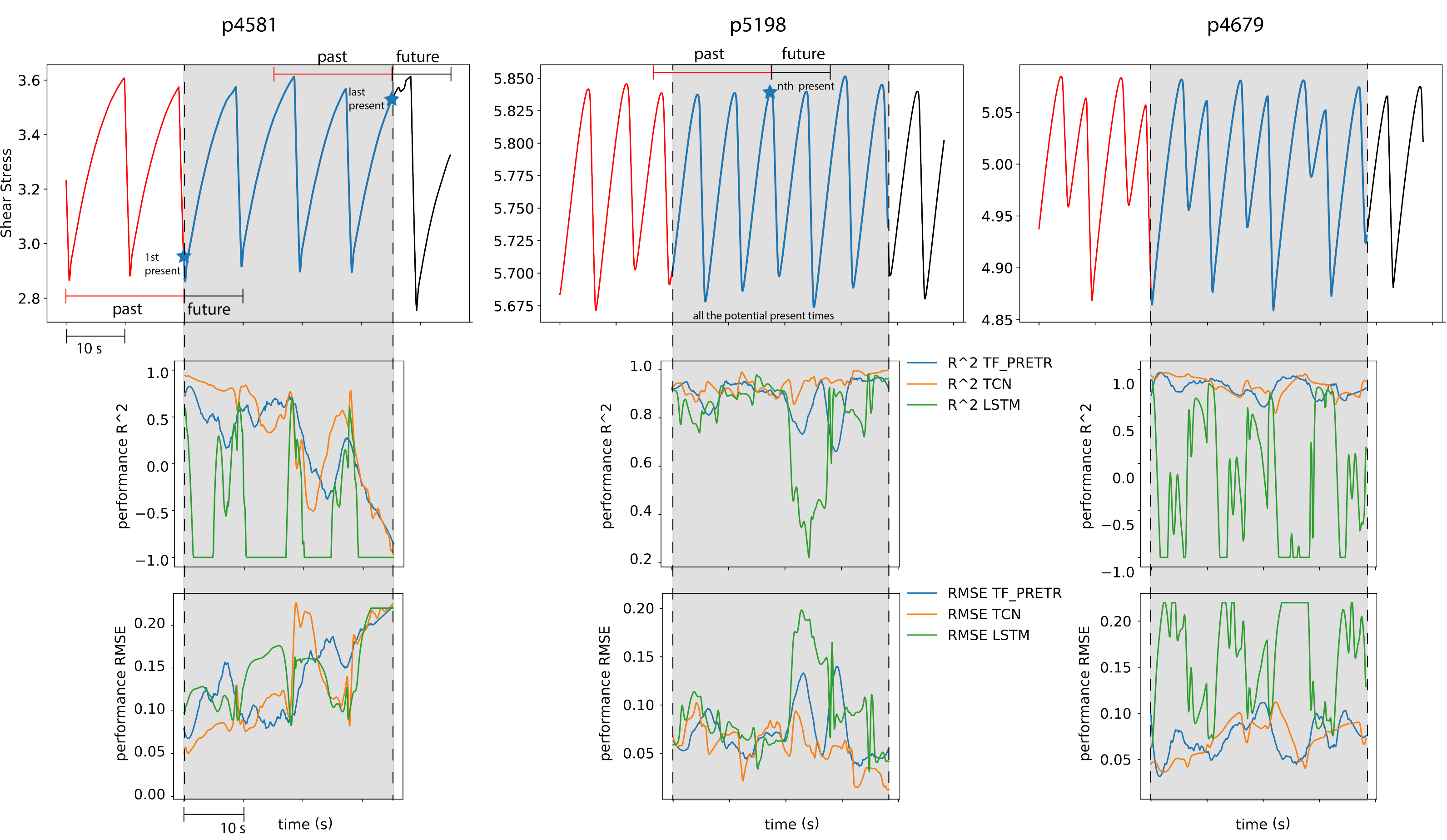}
    \caption{Results of the AR forecasting shown as performance variation with respect to present time.
    Red and black sections represent input for the firsts windows and ground truth for the lasts windows (respectively). Blue sections are times when model predictions are compared to data, with performance shown below (stars in Figure \ref{fig:pt2 AR performance}).
    Note that $R^2$ metric is not shown below $-1$ and RMSE is not shown above $0.22$. For TF we show a smoothed version of the original, noisier data because TF is the most complex model with the largest variation in performance (e.g., Gieger et al., 2020)}
    \label{fig:pt2 AR stress and metrics}
\end{figure}

\begin{figure}[H]
    \centering
    \includegraphics[width=0.4\textwidth]{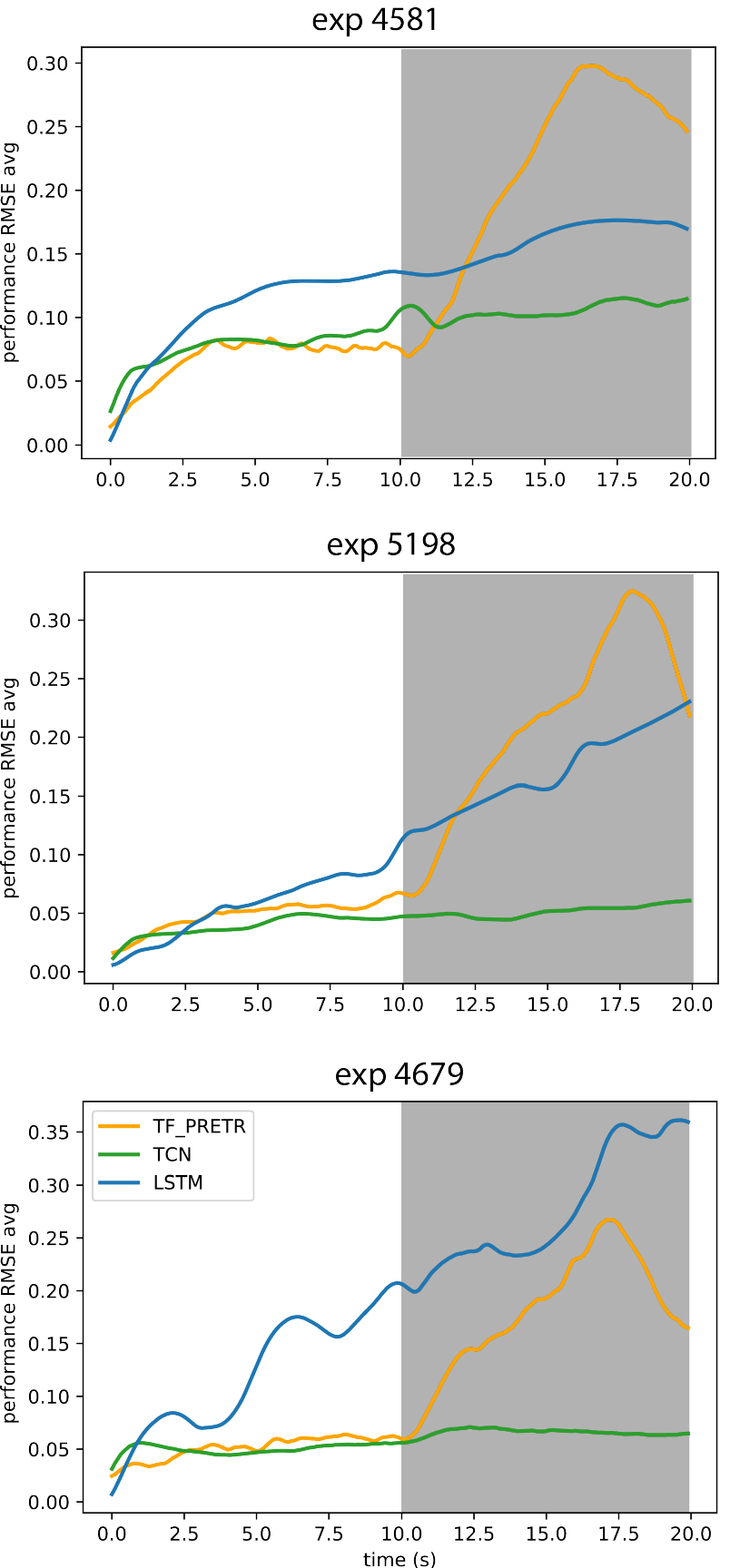}
    \caption{Data showing how AR model performance evolves using a 20 second window for forecasting. Note that the lab seismic cycle is variable but typically < 10 sec. (Figure 6). Performance values are the RMSE of the average of all the i-th at each time predictions compared with the i-th time ground truth value, in all the windows in the test set. We use only RMSE because it is not possible to compute $R^2$ point by point.
    White section is for times from present to present+10s (100 steps in the future). Gray section is for times from present+10s to present+20s (from the 100th to the 200th step in the future).
    The model has been trained to predict 100 steps in the future, so this is just a test to investigate the generalizability of the procedure.}
    \label{fig:pt2 performance_degradation_avg}
\end{figure}


\clearpage

\section{Supplementary material}
\beginsupplement

\subsection{Adopted Deep Neural Network architectures}
\label{s:Adopted Deep Neural Network architectures}
We formulate sequence modelling as a regression task, i.e. lean to minimize the Euclidean distance between the true and the predicted values.
Specifically, the sequence is a time series and the model needs to respect causality, i.e. in order to predict the output $y_t$ for some time $t$, we can only use those inputs that have been previously observed $(x_{t-k}, ..., x_{t-1})$. The problem faced in prediction (discussed in \ref{ssec:PART 1- Prediction model}) is supervised because given the input to the model, we know the output that it should produce and therefore we can compute gradient and train the network to turn out the right output. The forecasting explored in \ref{ssec:Forcasting} is also a regression problem, but learn self-supervisedly since we aim to predict the future from the past: the model needs to learn data representations to solve the task.

We will briefly present here the theory behind the architecture adopted for this work.

\subsubsection{1D-CNN}
Convolutional neural network (CNN) is a class of deep and supervised models that was introduced for the first time by LeCun et al. in 1998 for processing data that has a grid-like topology (e.g. images): this first CNN was applied to digit recognition, using MNIST dataset. CNNs get a dominant class of deep learning methods after the ImageNet competition for image recognition \citep{ref:Krizhevsky2012}, which has then become popular in the most varied applications. A typical architecture of a 2D convolutional network
consists of a set of layers each of which contains several filters for detecting various features in the input image, the model performs convolutions using the chosen kernels and in doing this the procedure adds activation functions (i.e. activation when a feature is matched), constituting the so called feature map.\\
With 1D-CNN we can do the same for a 1-Dimensional input, e.g. a temporal series. We have a 1-Dimensional array in input and some 1-Dimensional kernels that we use to perform convolution and extract features, as in the 2D case. This is indeed the main difference between 1D and 2D CNNs: 1D arrays replace 2D matrices for both kernels and feature maps. That result in a low computational complexity: $\mathcal{O}(k*n*d)$ for 1D CNNs, with respect to $\mathcal{O}(k*n*d^2)$ of 2D CNNs. Where $k$ is the kernel size of the convolution, $d$ is the representation dimension or embedding dimension of a word, $n$ is the sequence length.


\subsubsection{LSTM}
\label{ss:Long Short Term Memory (LSTM)}

In the past years, LSTM (Long Short-Term Memory network) \citep{ref:Hochreiter1997} has been successfully applied to a number of sequence model tasks, e.g. speech recognition, language modeling and translation, image captioning, trajectory forecasting and so on. In this work we apply it in geophysical problems.\\
LSTM it's a type of Recurrent Neural Network (RNN): RNNs are deep learning models that iteratively combines past informations with the present, to make them persist. Indeed, they have an “internal state” (hidden state) that can be seen as the memory: it is updated as a sequence is processed, by applying a recurrence formula at every time step, using function that combine the past information with the current input. In figure \ref{fig:RNN} left is represented the RNN if we unroll the loop.
In figure \ref{fig:RNN} right is represented one iteration of the RNN, where:

\begin{align*}
    h_t &= tanh(W_{hh}h_{t-1}+W_{xh}x_{t})=\\
    &= tanh(\begin{bmatrix}W_{hh}  W_{xh}\end{bmatrix}\begin{bmatrix} h_{t-1} \\ x_{t}\end{bmatrix})=\\
    &= tanh(W\begin{bmatrix} h_{t-1} \\ x_{t}\end{bmatrix})
\end{align*}

here $tanh()$ is the non-linear function, $W$ are the parameters, $h_t$ and $h_{t-1}$ are the hidden state at time $t$ and $t-1$ and $x_t$ is the input at time $t$.\\

LSTM works, for many tasks, much better than the RNN standard version. They were introduced in 1997 \citep{ref:Hochreiter1997}, and were improved and popularized by many people in following works. 
The main problem of the standard RNNs is the difficulty to access information from many steps back. LSTM instead is explicitly designed to avoid the long-term dependency problem: they are capable of learning long-term dependencies, thanks to some internal mechanisms, called gates, that can regulate the flow of information. These gates can learn which data in a sequence are important to keep or throw away. Another important feature of LSTM is the Cell $c_t$ that performs better in forward (direct connection with past element) and in backward (easy backward of the model and avoid gradient vanishing problem, that is a common problem of other RNNs).\\
Here there are two internal states: $c_t$ and $h_t$ that proceed in parallel, and represent respectively the long and the short term memory. There is a complex mechanism to manage memory, by using four gates: 
\begin{itemize}
    \item Input gate (i): whether to write to cell
    \item Forget gate (f): whether to erase cell
    \item Output gate (o): how much to reveal cell
    \item Gate gate (g): how much to write to cell
\end{itemize}

In Figure \ref{fig:unrolled RNN_LSTM_remake} is represented one iteration of the LSTM. Details are provided by Formula \ref{formula:LSTM}, where $tanh()$ and $\sigma=sigmoid()$ are the non-linear functions, $W$ are the parameters, $h_t$ and $h_{t-1}$ are the hidden state at time $t$ and $t-1$, $c_t$ and $c_{t-1}$ are the cell state at time $t$ and $t-1$ and $x_t$ is the input at time $t$. The "forget gate" say how much we should be forgetting about the previous cell information ($c_{t-1}$ is the memory of our system) and then, once decided what to forget we would be modulating with an "input gate" how much we want to memorize from the current input $x_t$.

\begin{align}
    & \begin{bmatrix}i \\ f \\ o \\ g\end{bmatrix} = \begin{bmatrix} \sigma \\ \sigma \\ \sigma \\ tanh \end{bmatrix}W\begin{bmatrix} h_{t-1} \\ x_{t}\end{bmatrix}
    \label{formula:LSTM}
\end{align}

\begin{align*}
    & where: \\
    & c_t=f \odot c_{t-1}+i \odot g \\
    & h_t=o \odot tanh(c_{t})
\end{align*} 
To better understand the behavior of the memory, let's assume we are at time t, then the LSTM memory explicitly consider all the information from time $t-k$ to $t$. The best length of the long term memory ($k$) is not known a priori: we further analyse it in the dedicated section below \ref{sss:Best length for the past memory of LSTM}.\\
Here the computational complexity is: $\mathcal{O}(n*d^2)$, where $d$ is the representation dimension or embedding dimension of a word, $n$ is the sequence length.


\subsubsection{Transformer Network}
\label{ss:Transformer network}
This model was introduced in 2017 by \cite{ref:Vaswani2017} and it was born mainly for common natural language processing, but nowadays is successfully used in a variety of different  sequence modeling tasks (e.g. video, audio and so on). It has an encoder-decoder structure where the encoder maps an input sequence of symbol representations ($x_1, ..., x_n$) to a sequence of continuous representations $\textbf{z} = (z_1, ..., z_n)$. The decoder uses \textbf{z} to generates autoregressively an output sequence ($y_1, ..., y_m$) of symbols.
The Transformer Network (TF) is implicitly autoregressive, in that we use the predicted output in the input of the next step (auto means that it feeds its own prediciton). In particular, in order to let the transformer deal with the input, this is embedded onto a higher D'-dimensional space using a linear projection with a matrix of weights. In the same way, the output is a D''-dimensional vector prediction, which is back-projected to the original 1-D space. \\
Differently from RNNs that receive one input at a time, TF receives all inputs one-shot. The TF uses a "positional encoding" to encode time for each past and future time instant $t$. Positional encoding is necessary to give an ordered context to the non-recurrent architecture of multi-head attention, because without it the model is permutation invariant. Sine/cosine functions are used to define positional encoding vector, that is: we represent the time in a sine/cosine basis.\\

The Transformer has 3 fundamental modules (attention, fully connected, residual connections). The attention modules are 2: self-attention and encoder-decoder attention. 
The encoder (Figure \ref{fig:transformer model Vaswani_remake}, left) has six identical layers, where each layer has two sub-layers: a  multi-head self-attention mechanism and a position-wise fully connected feed-forward network. All the outputs have a dimension of $d_{model} = 512$ . The decoder (Figure \ref{fig:transformer model Vaswani_remake}, right) has six identical layers and it has an additional layer in addition to the two sub-layers already described in the encoder: this performs multi-head attention over the output of the encoder stack. Decoder uses both self-attention and encoder-decoder attention, but the self-attention sub-layer in the decoder uses a masking mechanism to prevent positions from attending to subsequent positions. This ensures that the predictions for position $t_i$ can depend only on the known outputs at positions before $t_i$. To start forecasting it uses a special token that indicate the start of the sequence. It is shown with <S> in \ref{fig:transformer model Vaswani_remake}. The "Add \& Norm" layers in figure \ref{fig:transformer model Vaswani_remake} refer to Residual Connections (that sum the output of each layer with the input, to avoid vanishing gradient problem) and Layer Normalization.\\

The attention function maps a query and a set of key-value pairs to an output, where the query Q (dimension $d_N \times d_k$, where $d_N$ is the number of element in the sequence and $d_k$ the latent dimension), keys K (dimension $d_N \times d_k$), values V (dimension $d_N \times d_v$), and output are all vectors. Q is related with what we encode (it can be output of encoder layer or decoder layer); K is related with what we use as input to output; V is related with input, as a result of calculations, and it is a learned vector.
The output is computed as a weighted sum of the values, where the weight assigned to each value is computed by a compatibility function of the query with the corresponding key. 
\begin{equation}
Attention(Q,K,V)=softmax(\frac{QK^T}{\sqrt{d_k}})V
\end{equation}
Instead of performing a single attention function, the model linearly project the queries, keys and values h times with different, learned linear projections to $d_k$, $d_k$ and $d_v$ dimensions, respectively, then performing the attention function in parallel for each projected query, key and value. This allows the model to jointly attend to information from different representation subspaces at different positions: those are the heads, and we need more than one because each of these capture specific characteristic of the features.\\
For TF the computational complexity is: $\mathcal{O}(n^2*d)$, where $d$ is the representation dimension or embedding dimension of a word, $n$ is the sequence length.

\subsubsection{Transformer Network not pretrained forecasting results}
\label{sss:Transformer not pretrained forecasting results}

As explained in \ref{sssec:part 2 Transformer Network results}, TF is good in learning the aperiodicity and the singularities, however the common feature of all the experiments is the oscillatory behaviour of the signal. Without the pretraining with the sine wave, TF can't predict properly the target. \\
Moreover TF is the most complex among the tested model in optimization and training and it requires a lot of data and computing to start working. We have quite small dataset though, that are not enough in training properly the model. As shown in Table \ref{tab:TF comparison results pt2}, the results of TF not pretrained are always worst than the pretrained TF. Some windows of example for the three experiments are in Figure \ref{fig:pt2 AR performance TF not pretrained}.


\subsection{Networks training details}
\label{ss:Networks training details}
We train the model for 120 epochs in the case of prediction and for 30 epochs in the case of forecasting. In both cases we use the validation dataset to pick the best epoch and use it in testing phase, in order to avoid overfitting. 

The size of each batch is 256 or 32 for prediction or forecasting, respectively.

As optimizer we use NAdam in the case of prediction: this is like Adam optimizer with the difference that it uses Nesterov momentum. In the case of forecasting we use Noam: this is like Adam optimizer with the difference that it increases the learning rate linearly for the first steps, and decreases it after that proportionally to the inverse square root of the step number \citep{ref:Vaswani2017}.

Table \ref{tab:Training validation testing split} summarizes the number of data samples we have for each experiment, for training, validation and testing datasets. As explained in subsections \ref{sss:Prediction dataset split} and \ref{sss:Forecasting dataset split}, we adopt different dataset splits. This is because we use different frameworks (i.e. TensorFlow and PyTorch), which means different preimplemented functions. TensorFlow allows the user to select the validation part from the training data (so we take 10\% from the 70\% of the dataset used as training data). With PyTorch we can explicitly set train, val, test datasets sections (so we choose 70\%, 10\%, 20\%, respectively).
The reason why we use two different framework is that the model from LANL competition was in TensorFlow, then we keep this choice. Then we move to Pytorch for the forecasting part, since it's more straightforward and it makes model and functions editing easier.

\newpage

\subsection{Supplementary tables}
\begin{table}[ht]
\begin{center}
\begin{tabular}{|l|l|l|l|l|l|l|l|}
\hline
\multicolumn{2}{|l|}{}                     & \multicolumn{2}{l|}{Exp p4581} & \multicolumn{2}{l|}{Exp p5198} & \multicolumn{2}{l|}{Exp p4679} \\ \hline
\multicolumn{2}{|l|}{}                     & Length         & Shift         & Length         & Shift         & Length         & Shift         \\ \hline
\multicolumn{2}{|l|}{One-point prediction}  & 1.0            & 0.1           & 1.0            & 0.1           & 0.01           & 0.003         \\ \hline
\multicolumn{2}{|l|}{Sequence forecasting} & 1.0            & 0.1           & 1.0            & 0.1           & 1.0            & 0.1           \\ \hline
\end{tabular}
\end{center}
\caption{Data preprocessing is done using overlapped moving windows to calculate statistical features from the original data. Here, "Length" refers to time in seconds of the windows length and "Shift" is the time in seconds by which windows are shifted. Acoustic data are recorded at 4 MHz, thus a 1 s window with a 0.1 s shift means that we produce 10 statistical features per second.  We varied window size for each experiment and chose values that produced optimum results. 
"One-point prediction" refers to the first part of our work where we use LSTM+CNN model to predict one point at a time based on the prior signal variance. "Sequence forecasting" refers to the second part of the work where we use AR models (LSTM, TCN or TF) to forecast a sequence of values at future times in an auto-regressive fashion.}
\label{tab:win_len_shift}
\end{table}
\begin{table}[ht]
\begin{center}
\begin{tabular}{|c|c|c|c|c|}
    \hline
    Dataset & Task & p4581 & p5198 & p4679 \\
    \hline
    \multirow{2}{*}{Train} & Prediction & 1829 & 1832 & 38237 \\
    & Forecasting & 17300 & 18100  & 18000 \\
    \hline
    \multirow{2}{*}{Validation} & Prediction & 203 & 203  & 4248 \\
    & Forecasting & 1900 & 2100 & 2000 \\
    \hline
    \multirow{2}{*}{Test} & Prediction & 902 & 903 &  19066 \\
    & Forecasting & 3800 & 4100 & 4000 \\
    \hline
\end{tabular}
\end{center}
\caption{Training, validation and testing dataset sizes. For prediction this is the number of datapoints, for forecasting this is the number of windows (each window includes past-input and future-output)}
\label{tab:Training validation testing split}
\end{table}

\begin{table}[ht]
\begin{center}
\begin{tabular}{|c|c|c|c|c|}
    \hline
    \multirow{2}{*}{Target} & \multirow{2}{*}{$R^2$}  & p4581 & p5198 & p4679 \\
    &    & Glass beads & Quartz powder & Quartz powder \\
    \hline
    \multirow{2}{*}{Shear Stress} & LSTM+CNN & 0.9254 & 0.9884 & 0.9574 \\
    & XGBoost & 0.73 & 0.83  & --- \\
    \hline
    \multirow{2}{*}{Time To Start Of Failures} & LSTM+CNN & 0.6317 & 0.9313  & 0.8229 \\
    & XGBoost & --- & 0.85 & 0.70 \\
    \hline
    \multirow{2}{*}{Time To End Of Failures} & LSTM+CNN & 0.8721 & 0.9697 & 0.9200  \\
    & XGBoost & --- & --- & 0.86 \\
    \hline
\end{tabular}
\end{center}
\caption{Comparison between our results obtained with NN model vs. available results from the literature obtained with ML (XGBoost model) \citep{ref:Hulbert2019, ref:RouetLeduc2017}. For each target we show $R^2$, since RSME is not available from the literature. Our procedure outperforms the state-of-the-art in all the available occurrences.}
\label{tab:exp results pt1_VS_XGBoost}
\end{table}

\begin{table}[ht]
\begin{center}
\begin{tabular}{|c|c|c|c|c|}
    \hline
    \multirow{2}{*}{Model} & GOF & p4581 & p5198 & p4679\\
    & Material & Glass beads & Quartz powder & Quartz powder \\
    \hline
    \multirow{2}{*}{TF pretrained} & $R^2$ & 0.1172 & 0.8914 & 0.7940 \\
    & $RMSE$ & 0.1460 & 0.0707 & 0.0738\\
    \hline
    \multirow{2}{*}{TF not pretrained} & $R^2$ & $-$0.3410 & 0.6376 & 0.6061 \\
    & $RMSE$ & 0.1510 & 0.1247 & 0.0997\\
    \hline
\end{tabular}
\end{center}
\caption{Experimental results for autoregressive forecasting. This is a comparison between the TF models, when pretrained and when not. The goodness of fit (GOF) is an average computed among all the tested windows. Figures for TF pretrained, together with all the tested models are in Figure \ref{fig:pt2 AR performance}. Illustration for TF is in Figure \ref{fig:pt2 AR performance TF not pretrained}.}
\label{tab:TF comparison results pt2}
\end{table}

\newpage

\subsection{Supplementary Figures}
\begin{figure}[H]
    \centering
    \begin{subfigure}[b]{\textwidth}
        \includegraphics[width=\linewidth]{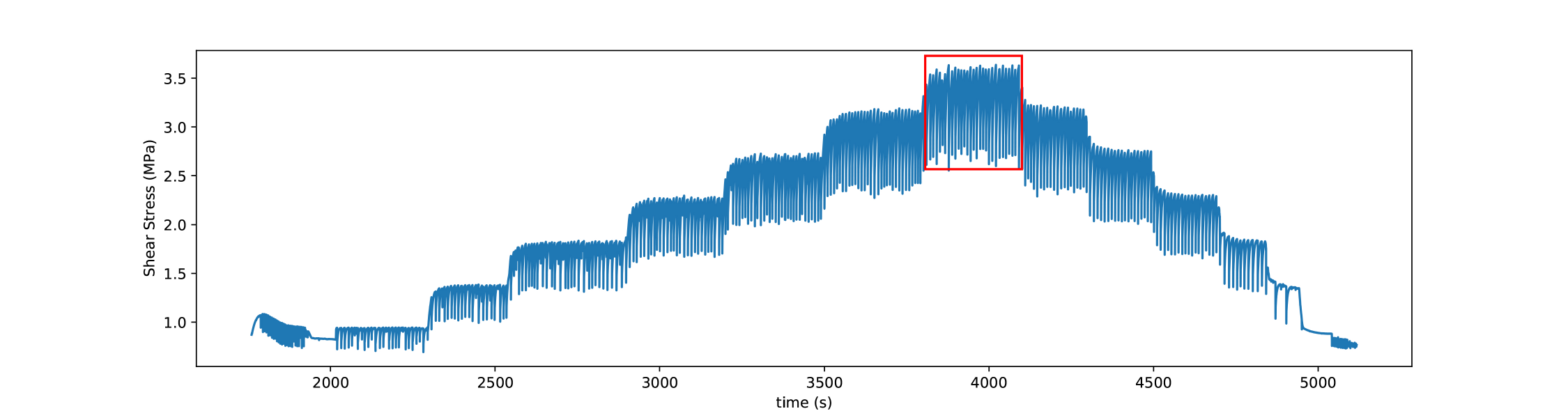}
        \caption{p4581}
        \label{fig:p4581_ALLexp}
    \end{subfigure}
    \begin{subfigure}[b]{\textwidth}
        \includegraphics[width=\linewidth]{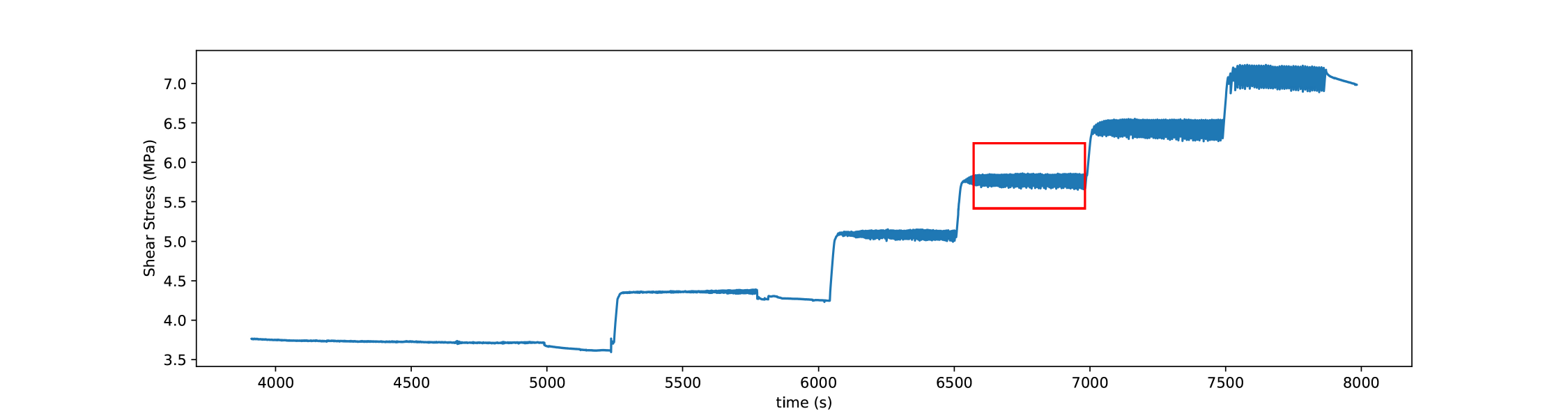}
        \caption{p5198}
        \label{fig:p5198_ALLexp}
    \end{subfigure}
    \begin{subfigure}[b]{\textwidth}
        \includegraphics[width=\linewidth]{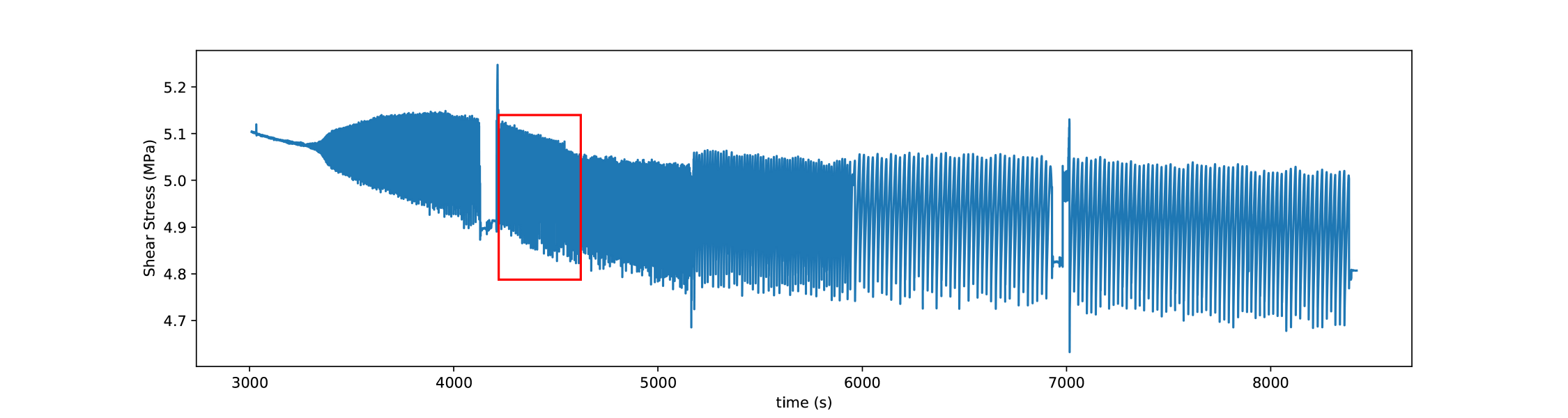}
        \caption{p4679}
        \label{fig:p4679_ALLexp}
    \end{subfigure}
    \caption{Full experiments, red box is for the subsection adopted in this work}
    \label{fig:ALLexp}
\end{figure}

\begin{figure}[H]
\centering
    \includegraphics[width=0.7\linewidth]{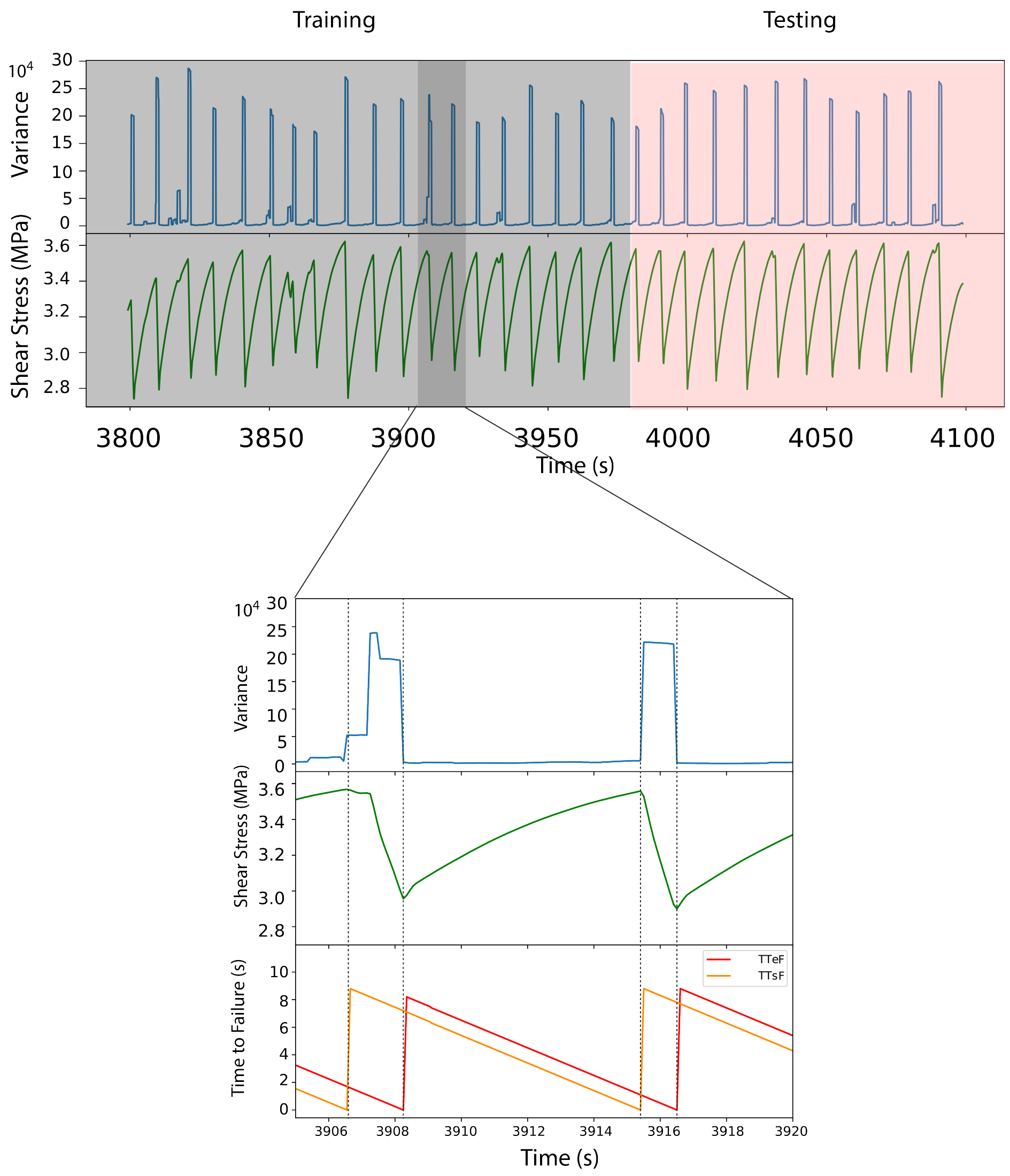}
    \caption{Signals' shape for experiment p4581, glass beads. For plot details see Figure \ref{fig:Signals' shapes}}
    \label{fig:Signals' shapes p4581}
\end{figure}

\begin{figure}[H]
\centering
    \centering
    \includegraphics[width=0.99\linewidth]{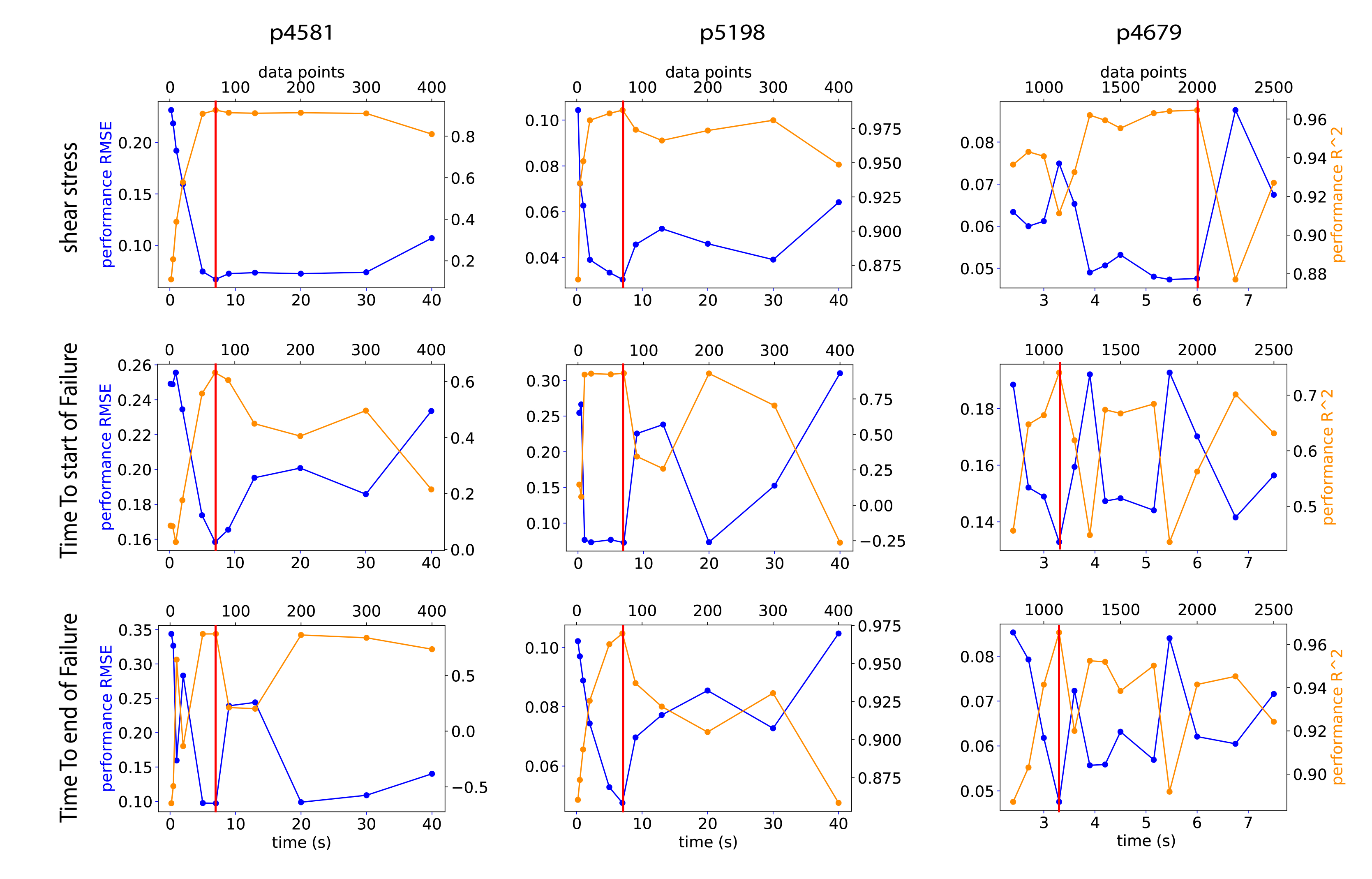}
    \caption{Variation in performance for different values of the LSTM memory length $k$. Each column shows results for one experiment.  Red line shows the optimum memory length time. For experiments p4581 and p5198 the optimum is about $k=70$, which  corresponds more or less to one seismic cycle. For experiment p4679 the optimum value is $k=2000$ when the target is shear stress while it is $k=1100$ when target is TTF. Here, one seismic cycle is about 1700 data points.}
    
    \label{fig:steps performances}
\end{figure}

\begin{figure}[H]
\centering
    \includegraphics[width=0.7\linewidth]{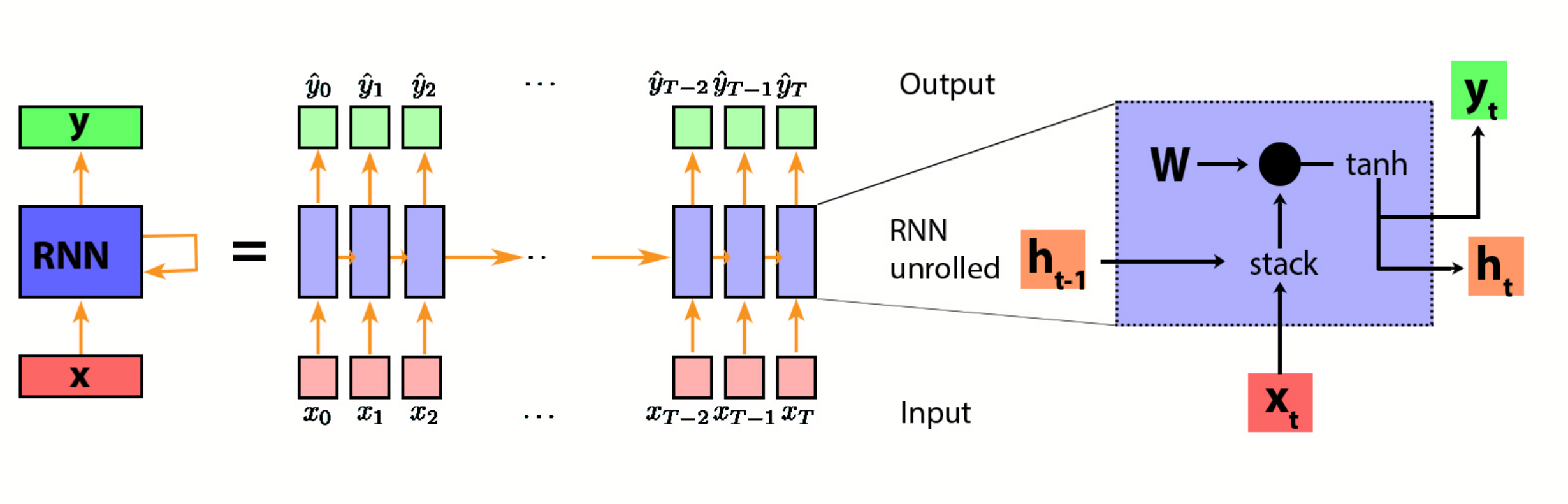}
    \caption{Recurrent neural network representation. The state-vector ($h_t$) is recursively updated at each t, including the new information coming from the current input.}
    \label{fig:RNN}
\end{figure}



\begin{figure}[H]
\centering
\includegraphics[width=0.9\linewidth]{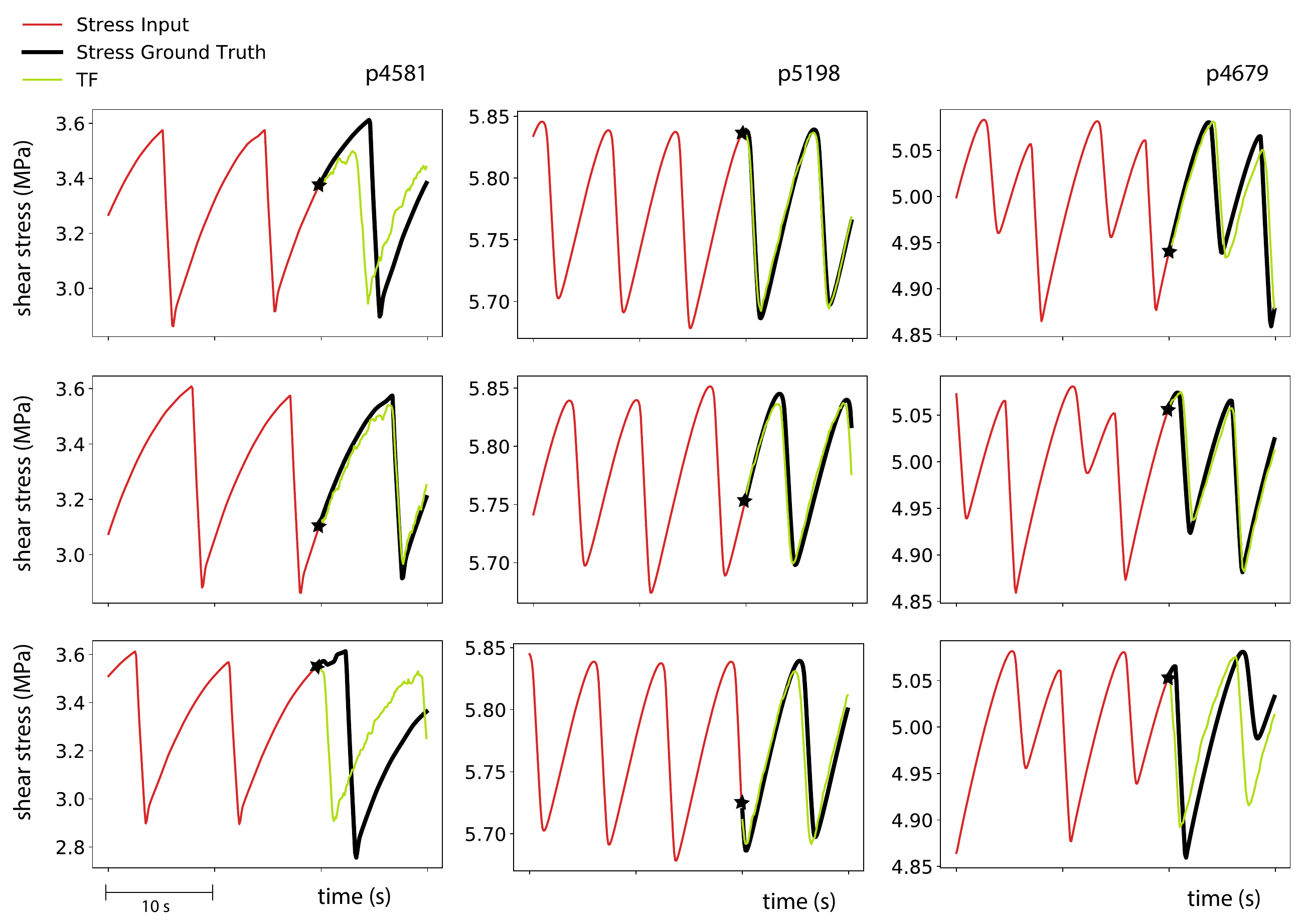}
    \caption{Results of forecasting models for the basic (not pretrained) Transformer Network.
    Each column shows a separate experiment. 
    Red lines show input data and black lines show ground truth data for forecast testing.  Green lines represent the output curves inferred from the model. X axis shows relative time, and Y axes are the target compared with model output. The results are not too bad for p5198 and p4679 and in general these results are worse than those for pretrained TF models.}
    \label{fig:pt2 AR performance TF not pretrained}
\end{figure}


\end{document}